# The impact of life-saving interventions on fertility

David Roodman[1]

April 2014

**Summary:** Many interventions in global health save lives. One criticism sometimes lobbed at these interventions invokes the spirit of Malthus. The good done, the charge goes, is offset by the harm of spreading the earth's limited resources more thinly: more people, and more misery per person. To the extent this holds, the net benefit of savings lives is lower than it appears at first. On the other hand, if lower mortality, especially in childhood, leads families to have fewer children, life-saving interventions could *reduce* population. This document critically reviews the evidence. It finds that the impact of life-saving interventions on fertility and population growth varies by context, and is rarely greater than 1:1. In places where lifetime births/woman has been converging to 2 or lower, saving one child's life should lead parents to avert a birth they would otherwise have. The impact of mortality drops on fertility will be nearly 1:1, so population growth will hardly change. In the increasingly exceptional locales where couples appear not to limit fertility much, such as Niger and Mali, the impact of saving a life on total births will be smaller, and may come about mainly through the biological channel of lactational amenorrhea. Here, mortality-drop-fertility-drop ratios of 1:0.5 and 1:0.33 appear more plausible. But in the long-term, it would be surprising if these few countries do not join the rest of the world in the transition to lower and more intentionally controlled fertility.

[1] Thanks to Alexander Berger for comments and guidance; and to Colin Rust for excellent and much-needed proofreading, as well as substantive commentary.

## Introduction

Many interventions in global health save lives. One criticism sometimes lobbed at these interventions invokes Thomas Malthus's famous theory of population. The good done, the charge goes, is offset by the harm of spreading the earth's limited resources more thinly: more people, and more misery per person. To the extent this holds, the net benefit of savings lives is lower than it appears at first.

But perhaps Malthus is wrong—at least today, at least in most places. Especially when children's lives are saved, couples may respond by having fewer children. This fertility reduction could partially or fully offset the effect of life-saving interventions on total population. In fact it could *more* than compensate, because parents may view children in part as investments in old-age security, and make those investments with great aversion to risk. Parents who are surer that each child will survive will become more certain *some* will be there for them later on, and will feel less need for the safety of "extra" children (Heer and Smith 1968, p. 106).

This document embodies an effort to review what is known about how mortality declines have affected fertility in poorer countries.

The starting question was the one just implied: Do parents compensate for the loss of children by having more or—to flip that—do they effectively compensate for additional lives saved by having fewer children? However, the confrontation with the evidence forced two conceptual refinements:

1. The question of the impact of mortality on fertility is abstract. There is no intervention that *only* saves lives. Distributing bed nets, for instance, prevents non-fatal as well as fatal cases of malaria. Modern programs for prevention of mother-to-child transmission (PMTCT) of HIV dispense drugs even as they encourage breastfeeding. This makes it hard to statistically distinguish the separate impacts of mortality, morbidity, and health advice. Inability to distinguish channels does make it harder to generalize from the available evidence to other diseases and interventions. But it is not a complete loss, since the practical question is often about the total impacts of a specific intervention such as bed net distribution, for which studies of the impact of malaria eradication are quite relevant. The difficulty of these distinctions also explains why our inquiry is more about the impact of *life-saving interventions* than that of *saving lives.* (But for focus, interventions that save few lives, such as deworming, are still excluded.)
2. Broadly, mortality and morbidity affect fertility through biological and volitional channels. A woman free of malaria is more likely to bring a pregnancy to completion in a live birth. A woman who loses an infant and thus stops breastfeeding is apt to begin menstruating again (cessation of lactational amenorrhea). These biological effects are distinct from parental decision-making (volition) in childbearing. This distinction too complicates the study of the impact of mortality on fertility. Some studies manage to make the distinction and some do not. All retain practical relevance.

As is typical in the social sciences, the phenomena we study are diverse while the evidence on them is fragmentary and suspect. This combination makes responsible generalization—estimation of the "truth"—quite hard. How couples in Sahelian northern Ghana decide whether to have another child is different from how couples in rural Bangladesh or urban America do. Culture, gender power dynamics, and economic circumstances all figure.[1] The impacts of an intervention on these dynamics depend on the specifics of the disease, the technology of the intervention, and its delivery.

[1] One dimension of difference is how much say the woman has in the decision to try for another child. For concision, I will sometimes speak of couples collectively "deciding" to have another child. But the power differences between man and woman should always be borne in mind.

But to make decisions, judgments must be made.

One fact beyond dispute is that fertility in developing countries, measured as lifetime births per woman, has fallen much faster than mortality since 1950 in developing countries. A "folk regression" of the fertility trend on the mortality trend would conclude that saving one life prevents two births: Malthus was completely wrong. Some studies reviewed here reach essentially that statistical result (though most are cautious about interpreting this correlation as causation).

I think the best interpretation of the available evidence is that the impact of life-saving interventions on fertility and population growth varies by context, above all with total fertility, and is rarely greater than 1:1. In places where lifetime births/woman has been converging to 2 or lower, family size is largely a conscious choice, made with an ideal family size in mind, and achieved in part by access to modern contraception. In those contexts, saving one child's life should lead parents to avert a birth they would otherwise have. The impact of mortality drops on fertility will be nearly 1:1, so population growth will hardly change.

In the increasingly exceptional locales where couples appear not to limit fertility much, such as Niger and Mali (Bongaarts 2013, p. 3), the impact of saving a life on total births will be smaller, and may come about mainly through the biological channel of lactational amenorrhea. Here, mortality-drop-fertility-drop ratios of 1:0.5 and 1:0.33 appear more plausible. Here, saving a life can be expected to increase population in the short-term. In the long-term, it would be surprising if these few countries do not join the rest of the world in the transition to lower and more intentionally controlled fertility.

After explaining some of the obstacles to the use of statistics for studying the impact of mortality on fertility, the review examines five kinds of evidence: historical, modern cross-country, modern cross-country panels, quasi-experimental, and large-sample microdata studies. A table summarizing my interpretations of the studies is in the conclusion.

## 1 Mortality and fertility: Trends and causes

Since 1950, humanity has progressed remarkably in lowering death and birth rates. The reasons for the declines in mortality are well known; prominent among them are advances in medicine and public health in combating infectious diseases, as well as global campaigns to deliver those advances. Figure 1 shows the trends by continent in child mortality, here defined as death before age 15 (UN Population Division 2013).

Fertility has also fallen dramatically—see Figure 2. The reasons behind this trend are broadly understood too. Reliable forms of contraception have been developed and made widely available. Economic growth has created opportunities for women to earn more outside the home, thus a reward for having fewer children. Earning power has given many women more voice within the household, including in matters of sex and contraception. Women's greater access to education has amplified the effects of economic growth by multiplying their earning potential and empowering them with greater knowledge of contraception. By the same token, rising access to and value of education has probably led parents to have fewer children while investing more in each of them, notably in their schooling. Finally, norms about family size have been shifted by public campaigns—forcibly in China, voluntarily most other places—and even by soap operas (La Ferrara, Chong, and Duryea 2012).

(Note that mortality rates and lifetime births/woman are statistical abstractions. For example, the figure of 5.83 births/woman in Asia for 1950–55 represents what would happen to a woman who spent all her fertile years in that hypothetically fixed context rather than in the actual context of falling fertility over subsequent years. Similarly the mortality figures represent the chance of death for the imaginary child who grows up entirely within a given five-year period, during which the health regime is held fixed.)

**Figure 1. Fraction of children dying before age 15**

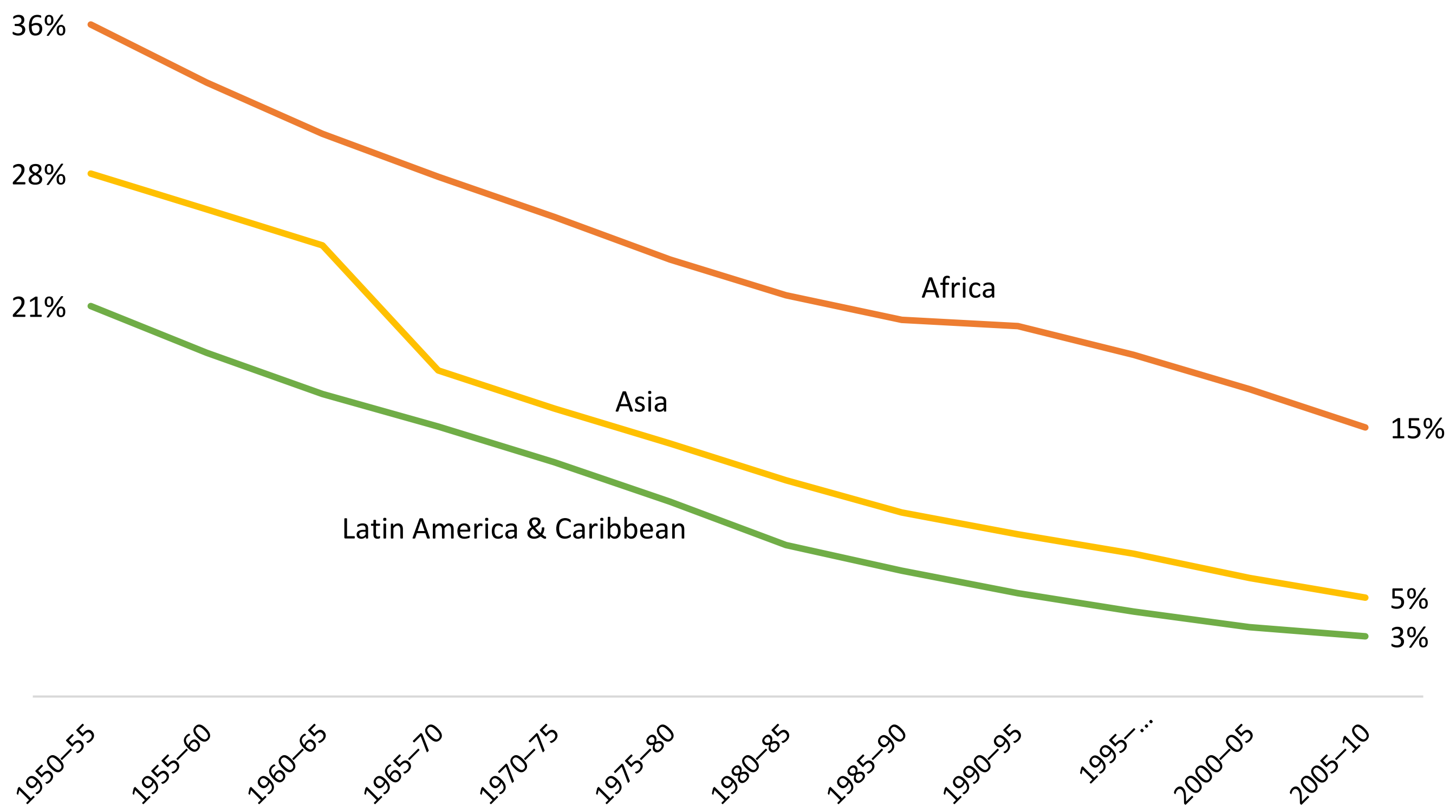

**Figure 2. Lifetime births/woman**

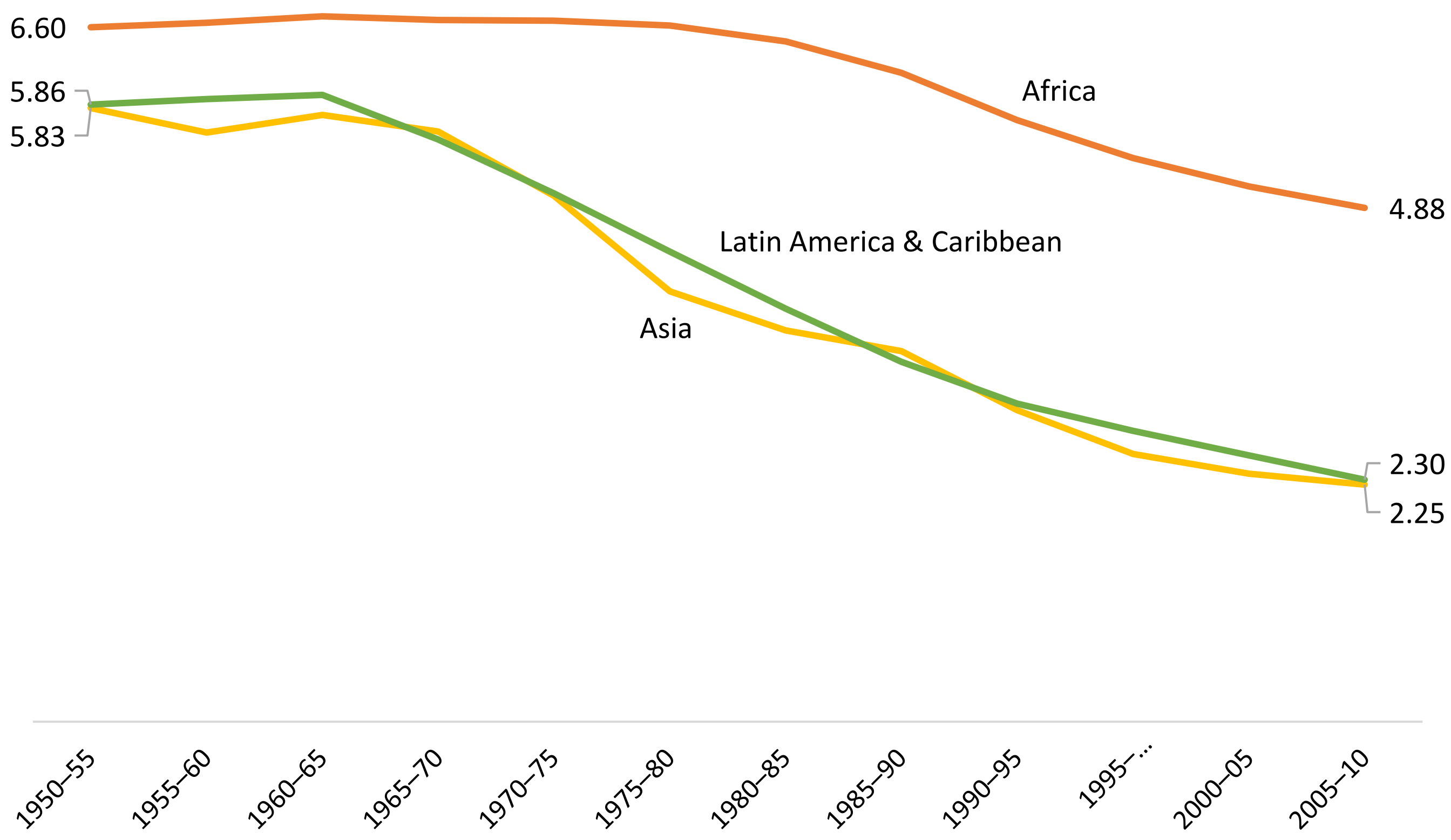

Mortality and fertility declines have exhibited the following broad patterns, most of which can be glimpsed in the graphs above:

- Mortality has declined steadily almost everywhere, the most globally significant exceptions being the disruption caused by the Great Leap Forward in China in the early 1960s and the stalling of progress in Africa circa 1990 with the spread of HIV.
- In contrast, in most countries, fertility initially held steady around traditional levels of 6 or more lifetime births/woman, then bent discernibly downward at a particular historical moment. Demographers speak of "onsets of fertility decline" as distinct phenomena. Today, only a few countries appear not to have reached such an onset.[2]
- Countries are reaching given mortality and fertility rates at lower GDP/capita levels than was once the case. Sub-Saharan Africa reached an infant mortality rate (deaths before age 1) of 7.9% in 2005–10 (UN Population Division 2013), when its GDP/capita was about $2,000 on a purchasing power basis (World Bank 2014, series "GDP per capita, PPP (constant 2005 international $)," region "Sub-Saharan Africa (all income levels)," year 2007). The United States achieved that mortality rate in 1921 (Bureau of the Census 1939, p. 23), by which time its GDP/capita in today's dollars had surpassed $5,000 (Bolt and van Zanden 2013).
- The same goes for onsets of fertility decline: although they arrive later in poorer countries, the more-recent onsets are occurring at lower levels of GDP/capita (and average education levels). Poor countries haven't had to become as rich (and educated) as they once did for fertility to begin falling (Bongaarts 2013, p. 7).
- These correlations are hardly mechanical. For example, some countries have reached the fertility decline onset at especially low levels of development while others entered it at higher levels. One reason appears to be large-scale campaigns to promote family planning. Kenya's fertility rate fell from 7.6 to 5.0 births/woman between 1975–80 and 1995–2000 while nearby Uganda's hardly budged, going from 7.1 to 6.7 (UN Population Division 2013). It is probably not a coincidence that Kenya was one of the first African countries to develop a population policy and that its fertility decline slowed after the mid-1990s, when funding for the program was cut (Bongaarts et al. 2012, p. 39–40). A similar comparison holds for Bangladesh and Pakistan: Bangladesh's much more rapid fertility drop "can plausibly be attributed" to its stronger family planning program (Bongaarts et al. 2012, p. 39).

From these broad patterns, we can infer that:

- Modern mortality and fertility declines are historically unprecedented in that never before have such poor societies so limited deaths and births.
- It is only in some of the very poorest countries, such as Mali, Niger, and Afghanistan, that mortality has fallen while fertility remains high, at 6 or more births/woman. It is there that population growth from reproduction (as distinct from immigration) is fastest.[3] Niger's population has tripled since 1975 for instance, and will triple again by 2038 under the UN's medium-fertility projection, which. The same scenario has Africa's total population expanding from 1 billion today to 4 billion in 2100 (UN Population Division 2013).
- While the mortality and fertility declines are broadly related, mortality declines are not the sole reason for fertility declines. One reason for this belief is somewhat evident in the graphs: it would be hard to explain why fertility held level for decades in many countries even as mortality fell. In particular, it appears that fertility onsets happen when the fraction of families engaging in family

[2] Total fertility and under-five mortality are graphed by country here, using the same UN data source.
[3] Qatar, Bahrain, the United Arab Emirates, and Kuwait had four of the six fastest-growing national populations in the 2000's, but mainly because of immigration. The rest of the top 20 were quite poor (UN Population Division 2013).

planning—deliberately limiting fertility—begins to rise. This decision is a function of many factors, of which the probability of losing a child is but one.

The last point is relevant to the question of how mortality affects fertility. Where couples do not limit fertility, mortality can only affect fertility through biological, not volitional, channels. Essentially, couples continue to have children as nature allows. At the opposite extreme, where couples aim for a specific number of children, such as two, then the volitional channel will be strong: when a couple loses a child, it is apt to replace it in pursuit of its desired family size.

All that is rather coarse generalizing, though. How mortality influences fertility almost certainly depends on many factors and has varied over time and place. So the bulk of this document reviews empirical studies meant to give us a sharper understanding of the link.

To help us think about implications for population growth, Figure 3 presents the mortality and fertility trends in a way that makes them more directly comparable, by estimating the number of child *deaths* per woman. As shown, in developing countries as a group, lifetime births per woman fell from 6.1 to 2.7. Meanwhile, the chance of a child in a developing country dying before age 1 plummeted from 15% to 5%; before age 5, from 25% to 7%; and before age 15 from 29% to 8%. Multiplying these percentages against total births, we find that the statistically average woman having 6.1 children in 1950–55 lost 1.8 of them before they grew up, taking age 15 as threshold to adulthood, while her counterpart in 2005–10 lost only 0.21 out of 2.7. Child deaths/woman fell by 1.6 while births fell by 3.4, for a ratio of about 1:2. ).

**FIGURE 3. LIFETIME BIRTHS AND CHILD DEATHS PER WOMAN, DEVELOPING COUNTRIES**

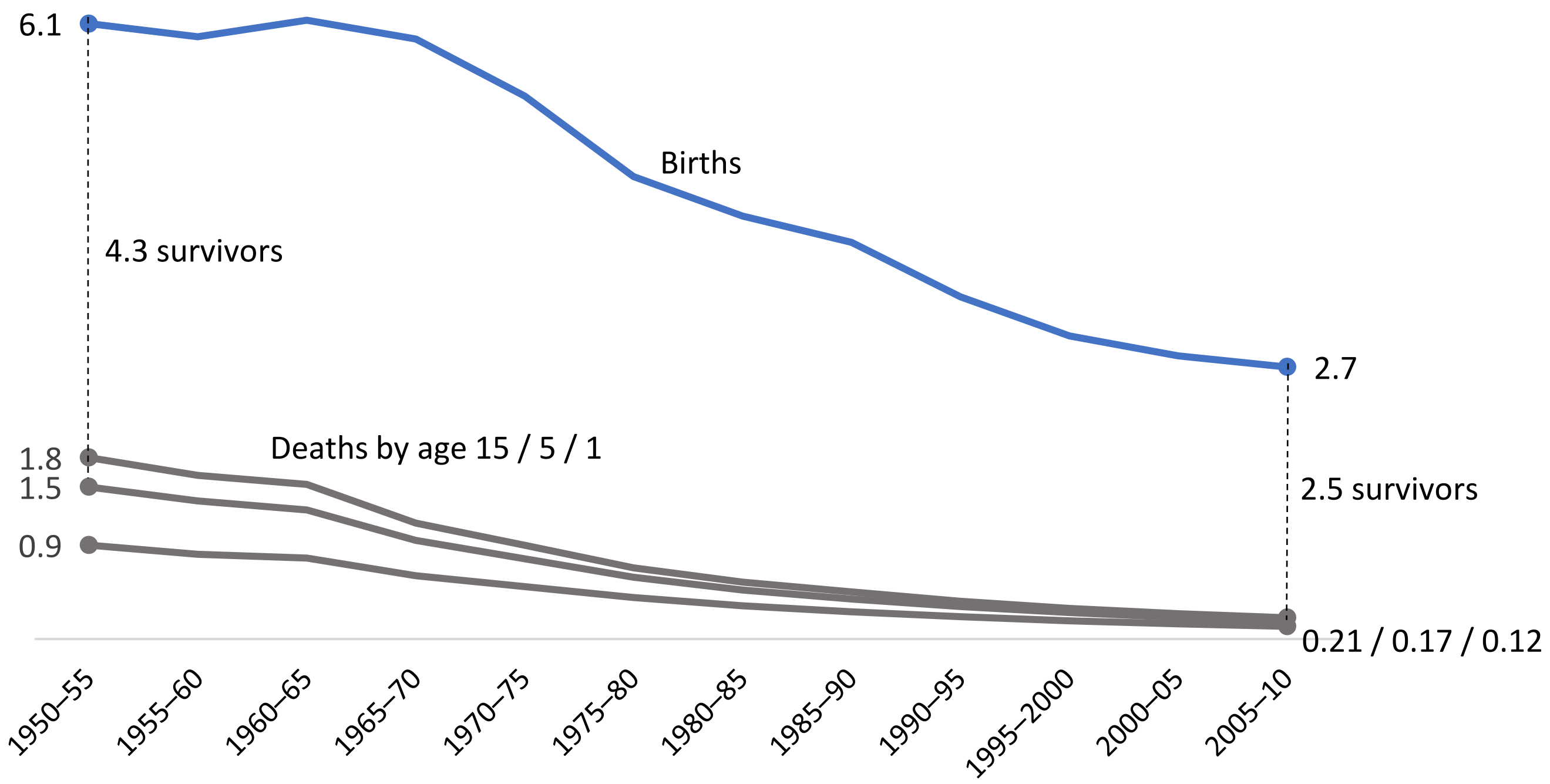

If we switch from tracking mortality and fertility over time to taking a snapshot of all countries at a specific time, the fertility-drop-mortality-drop ratio appears even larger. Figure 4 uses data for 1990, a year in or near the time periods of all the cross-country statistical studies examined below. For each country, it plots lifetime births/woman and lifetime under-five child deaths. Moving down the orange best-fit line from 8 to 4 births/woman (marked on the vertical axis) corresponds to a move from about 1.5 to 0.5 lifetime under-five deaths/woman (on the horizontal axis). That's a mortality-drop-fertility-drop ratio of 1:4. This cross-country pattern seems hard to fully explain as mortality leading to reduced fertility: would a family have four extra children if it lost one? But it is easy to understand as fertility affecting mortality, given how mortality is measured here: fewer children born to each woman means fewer who can die. This suggests that cross-country comparisons are particularly prone to overestimate the impact of mortality declines on fertility.

**FIGURE 4. TOTAL FERTILITY VS CHILD MORTALITY BY COUNTRY, 1990**

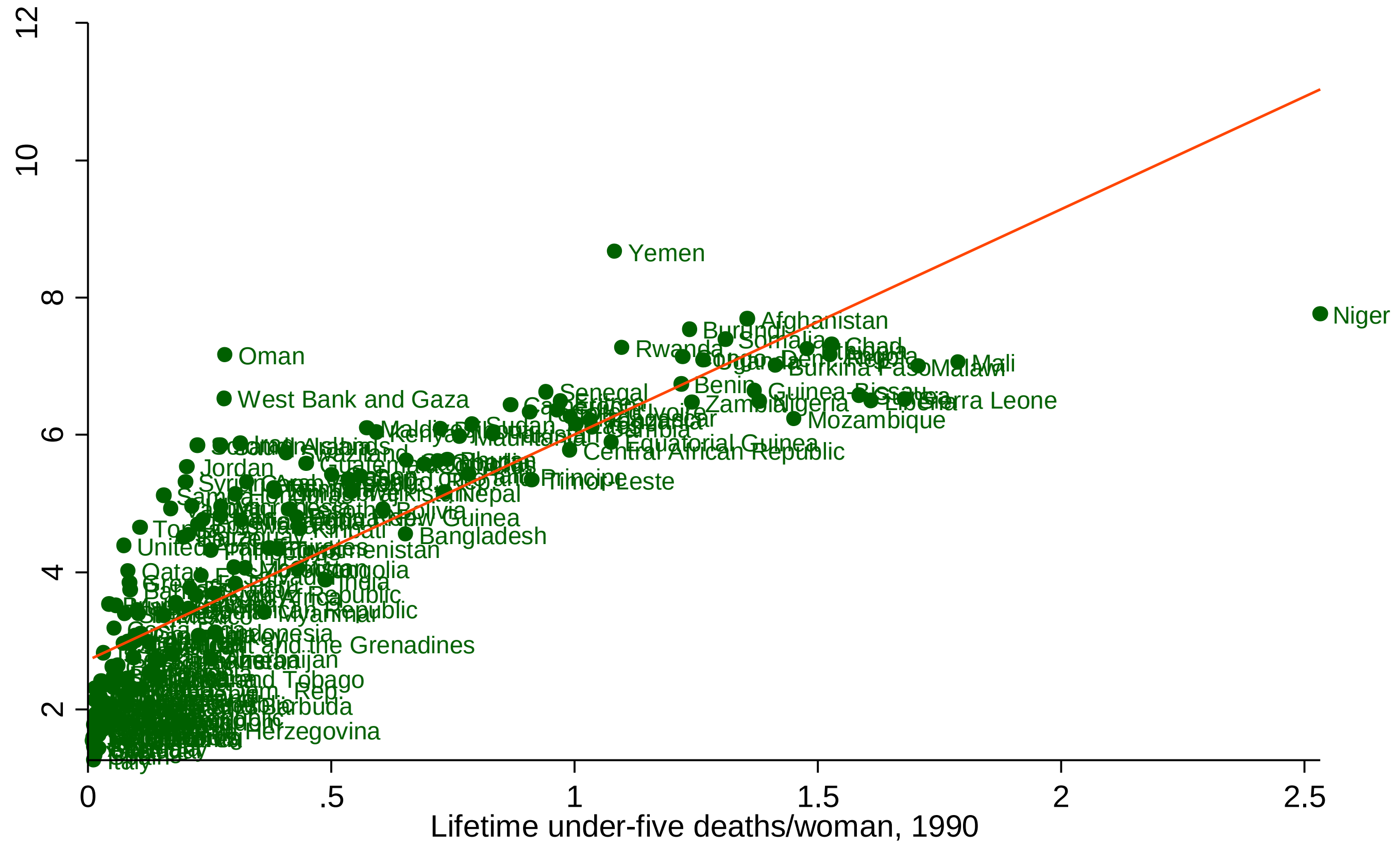

## 2 On the econometric challenges of studying causality

If we naively interpreted the mortality-fertility correlations in the graphs above as proving that mortality decline over the last 60 years caused *all* the fertility decline, we would conclude that the population increase from saving children's lives was offset 2 –4 times over by the consequent drop in fertility. Saving lives *reduced* population growth.

Of course, the story is more complicated than that. As already discussed, many factors are at work behind the two broad trends, some shared, some distinctive. And while mortality may itself affect fertility, causality also runs the other way. In developing countries overall, the chance of a baby surviving falls with the age of the mother (Rutstein and Winter 2014, p. 38, Table 13.). So reducing fertility early in womanhood also reduces mortality. Having fewer children in a family of limited means may also reduce effective competition for food and other necessities, making children less vulnerable to disease.

Clearly, fertility and mortality are connected by many causal arrows. It is the job of social science to ferret out the particular causal pathways we are interested in, from mortality to fertility. This section explains why social science has a hard job. Along the way, it introduces some terms and a visual language that will be used in discussing studies. For concreteness, it refers to Schultz (1997), the first study reviewed below.

The study gathers and analyzes data on national mortality and fertility rates for about 70 developing countries in 1972, 1982, and 1988/89 (Schultz 1997, Pg. 397). Among the causal relationships analyzed is the one of interest to us:

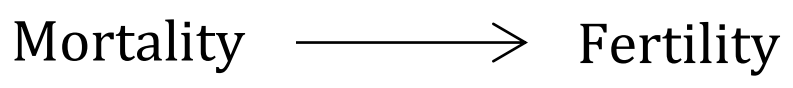

If this were known to be the only possible connection between mortality and fertility, then we could interpret any statistical correlation between them as measuring the sign and strength of that connection. But Schultz understands that the world works more like this:

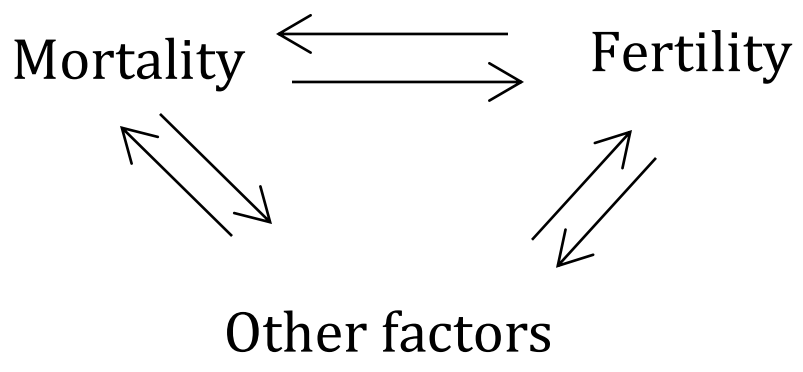

"Other factors" includes, for example, female education and GDP/capita, which could simultaneously cut deaths and births. They are "confounders" because their potential importance confounds any naïve attempt to attribute the mortality-fertility correlation to a simple causal story.

In the event, Schultz (1997), p. 398, Table 5, col 3, finds a correlation between mortality and fertility that is hard to ascribe to chance. Three families of theories could explain that, which I indicate by bolding arrows[4]:

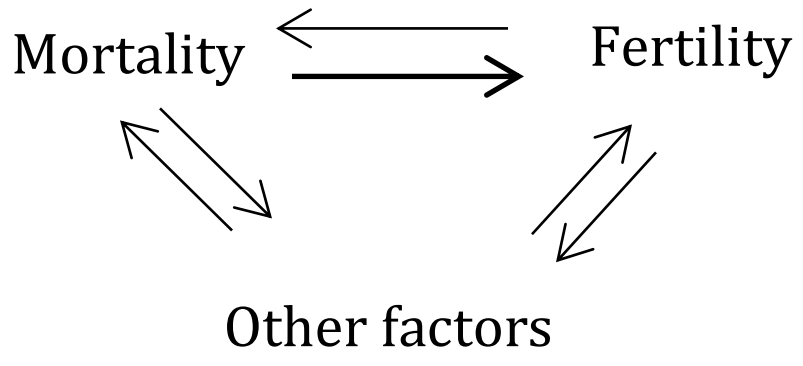

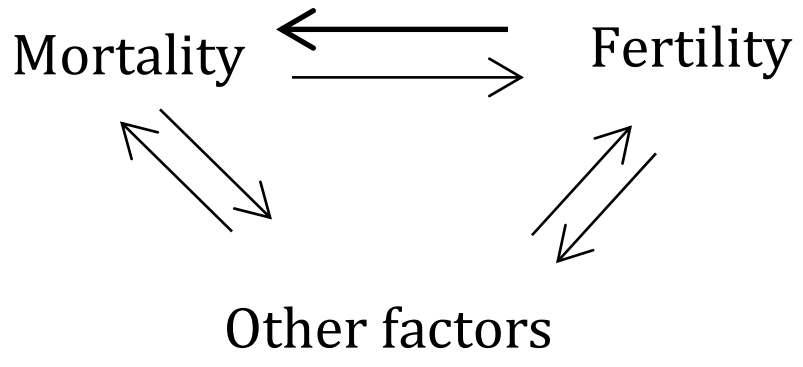

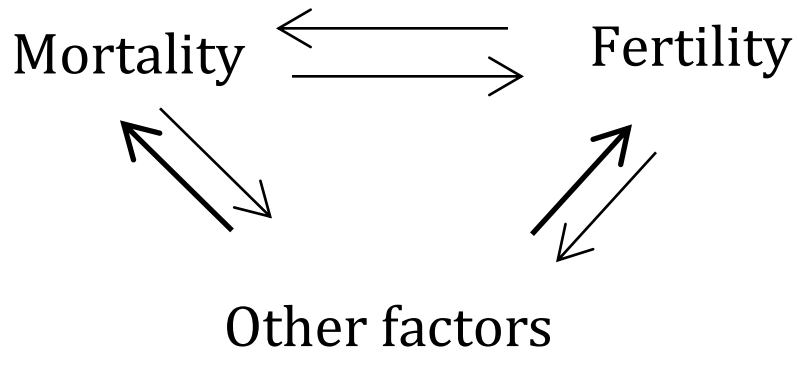

Schultz (1997) deploys two standard strategies to reduce the plausibility of the second and third theories. First, it "controls for" some of those other factors, such as GDP/capita. Roughly speaking, in instances where fertility and mortality fall just as GDP/capita rises, those drops are discarded as possible examples of the third theory above. In fact, Schultz (1997) controls for GDP/capita; education levels of men and of women, measured as average number of years of schooling in the current population; share of the labor force in agriculture; and the Muslim, Catholic, and Protestant shares of the population. Statistical analyses of this sort are called "regressions." (The controls are entered linearly, so nonlinear relationships between them and the variables of primary interest are not removed.)

[4] There is some conceptual redundancy in these graphs. Mortality might affect fertility *via* "other factors." But there are always intermediating factors. Evens neurons firing cause hands to move via intermediating mechanisms. So this possibility is indistinguishable from the first graphed.

But, the Schultz (1997) control set does not exhaust the possibilities for hidden third variables that could, behind the statistical scenes, reduce mortality and fertility at once and create the false appearance of causality from one to the other. Countries with better-run governments might have better public health programs, which reduce deaths and births in many ways.

This motivates the second strategy, which is to "instrument": Schultz (1997) introduces one more variable, national calorie intake/capita, along with an important assumption about its relationship with other variables. The picture looks like this:

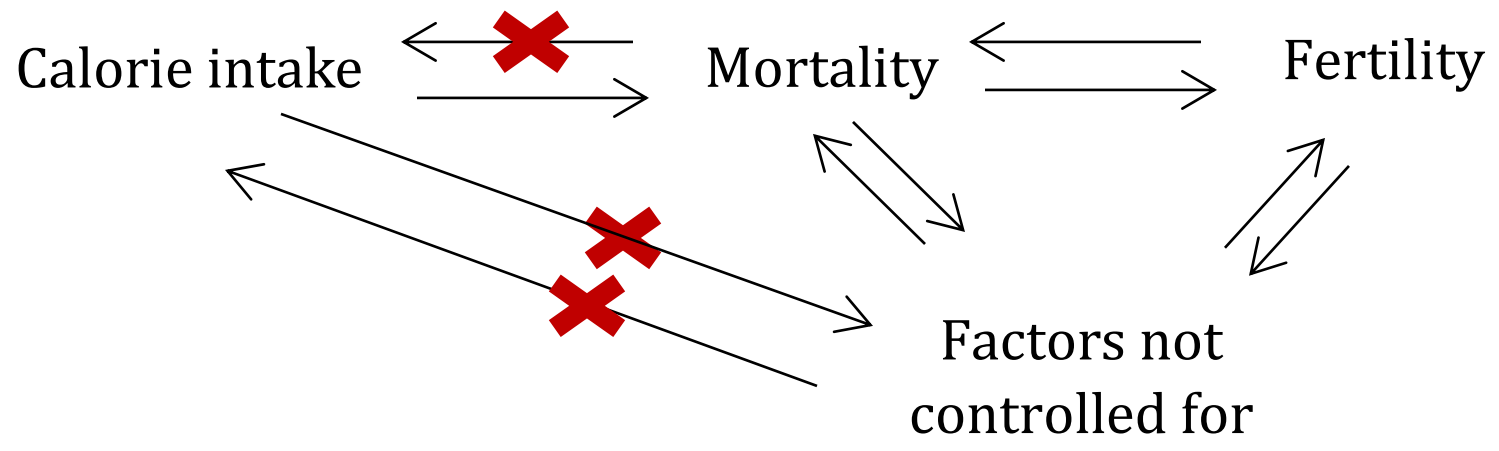

The red X's indicate causal pathways that are assumed *not* to operate; they are "exclusion restrictions." Calorie intake is said to be "exogenous" because it sits largely outside the causally entangled complex of other variables. One arrow that *is* assumed to operate certainly makes sense: calorie intake is believed to affect mortality. But according to the red X's, calorie intake does not affect fertility so directly. This assumption "is justified by biological and demographic investigations which conclude the effects of nutrition on reproductive potential or fecundity are negligible" (Schultz 1997, p. 400). To understand the power of this assumption, consider that Schultz (1997, p. 398, Table 5, col 4), detects a correlation between calorie intake and fertility. Within the confines of that diagram, only one theory can explain this link, indicated by the bolded arrows:

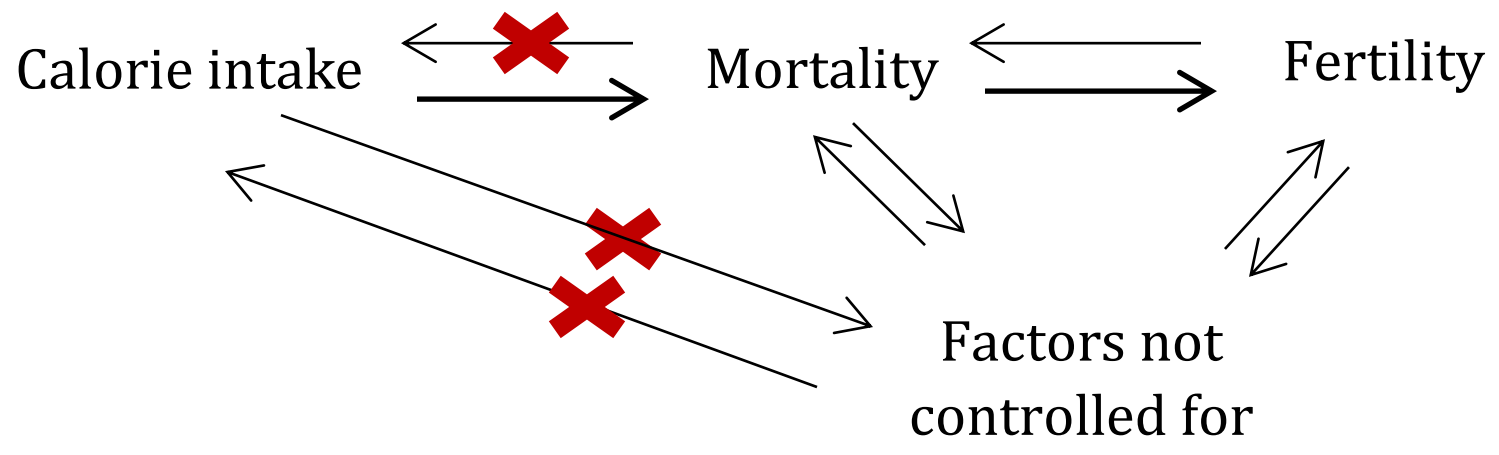

In particular, mortality affects fertility. Notice how the introduction of calorie intake into the analysis, along with assumptions about its causal relationship to other variables, lets Schultz (1997) study the impact of mortality on fertility. Calorie intake is said to "instrument for" mortality.

It's worth noting that the assumptions required to interpret the Schultz (1997) as the impact of mortality on fertility are actually stronger than Schultz suggests in citing "biological and demographic investigations." Those investigations only justify one of the three red X's depicted above, the one on the arrow from calorie intake to "factors not controlled for." (This prevents calories from affecting fertility by leapfrogging mortality.) If this were the only exclusion restriction we imposed, the world could work like this:

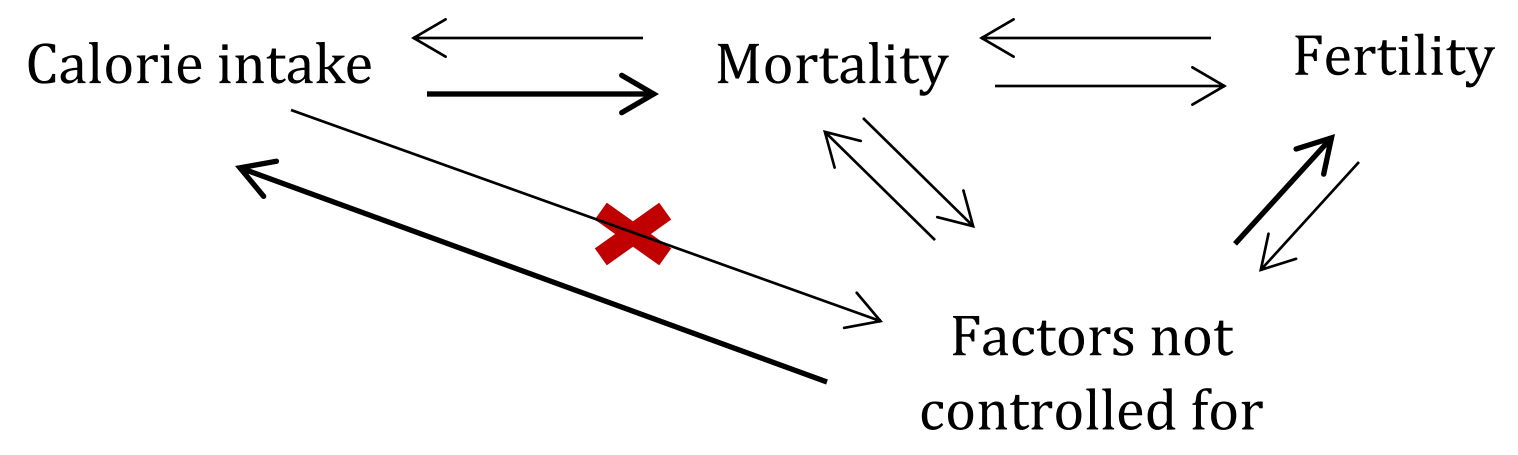

In words, a factor not controlled for might simultaneously affect fertility and calorie intake, the latter in turn affecting mortality. Perhaps economic inequality or government effectiveness accelerates progress on fertility out of proportion to its effects on mortality. Once again this would create a correlation between the variables of interest without any impact from one to the other.

Or conceivably the world works like this:

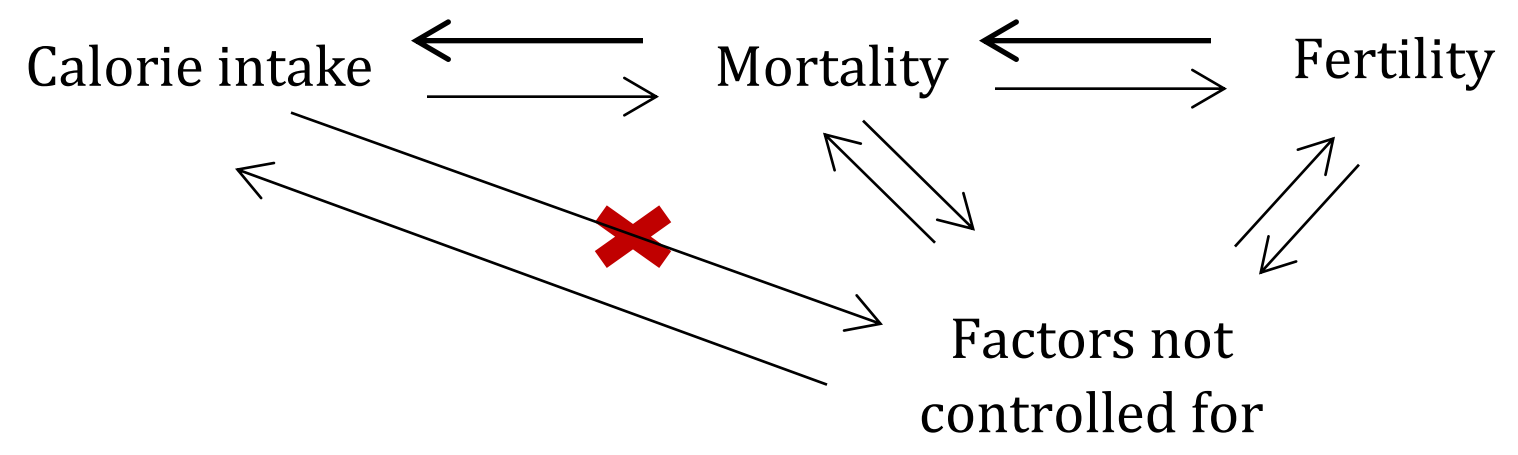

This shows that the technique of instrumenting, while useful, is not a cure all—and in fact works less well than many econometricians seem to recognize.

Lest you despair about the capacity of statistics to enlighten, imagine changing the picture in one way. Imagine a government tests a new, potentially lifesaving vaccine for children, but randomizes which villages receive it in order to study effects on mortality and knock-on effects to fertility. That would look like this:

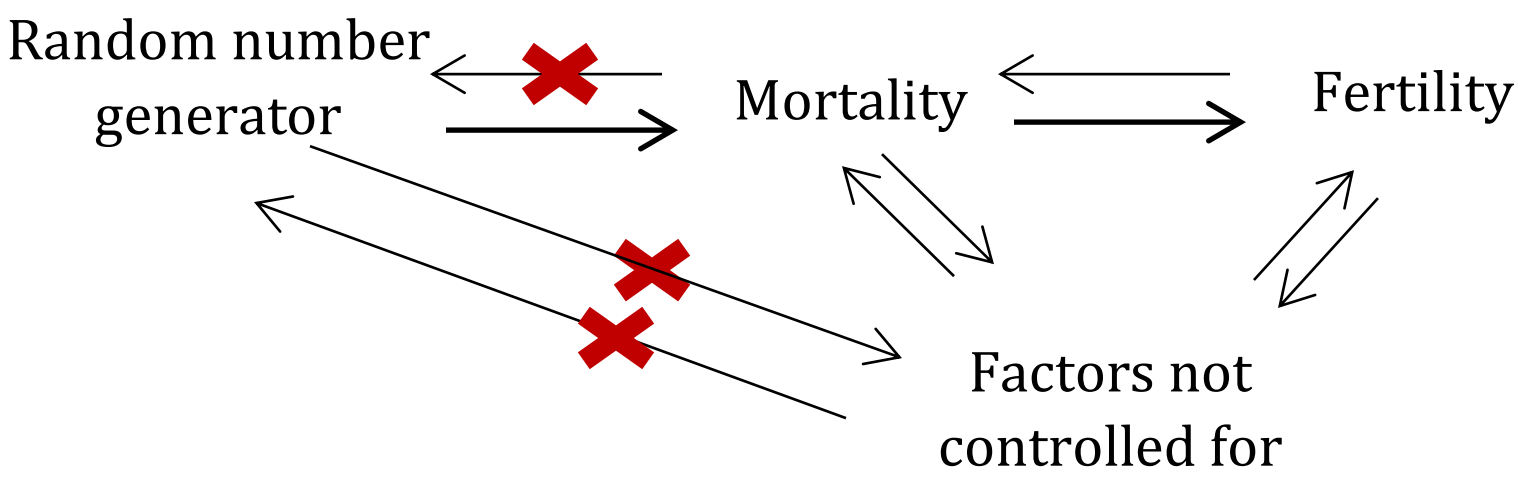

Now it is easy to believe all the red X's. No factor but chance would affect the random assignment of vaccines to villages. And randomly vaccinating some kids should not affect the fertility of their mothers. So if fertility eventually dropped more in villages receiving the vaccine, the case would be strong that lower mortality was an intermediate cause.

Absent randomization, interpreting statistical correlations as evidence of causation requires significant assumptions like those in the Schultz (1997) example above. This raises a fundamental question: if we are prepared to make such a broad assumption about how, e.g., calorie intake relates to other variables, why don't we just assume that mortality affects fertility and skip the econometrics? Putting that constructively, for a study to be useful, "the assumptions on which it rests must be more credible than the assumptions that it tests" (Roodman 2009a).

None of the studies in this review is randomized. However, they all incorporate steps to combat the sorts of issues raised above. Some control for lots of third factors. Others rely on time's arrow, arguing that if a mortality drop is *followed by* a fertility drop, the first is most likely to have caused the second. And some attempt to be “quasi-experimental” by exploiting an arbitrary event such as the sudden eradication of malaria from a country. A major focus of the review will be how well the various studies approximate the ideal of a controlled experiment, requiring only the weak assumptions in order to interpret their findings as causation from mortality to fertility.

## 3 Historical evidence

Quantitative historical economists (“cliometricians”) have harvested data from ancient records in order to estimate birth and death rates in northern Europe over centuries. Galor (2012) summarizes some of this work in a pair of graphs. They show deaths and births per year per thousand of population, here copied as Figure 5 and Figure 6. Death rates fell fairly steadily throughout the period, if apparently punctuated by outbreaks of disease and famine. As for fertility, except in France, it mostly declined only starting in the late 19th century.[5] That declining mortality preceded declining fertility suggests that the first caused the second. On other hand, the lags were so long—more than 150 years in England—that it is hard to put much faith in this interpretation. Galor (2012, p. 7–8), emphasizes the latter view, citing various quantitative analyses of such historical data.

**FIGURE 5. DEATH RATE OVER TIME IN EUROPEAN COUNTRIES, FROM GALOR (2012)**

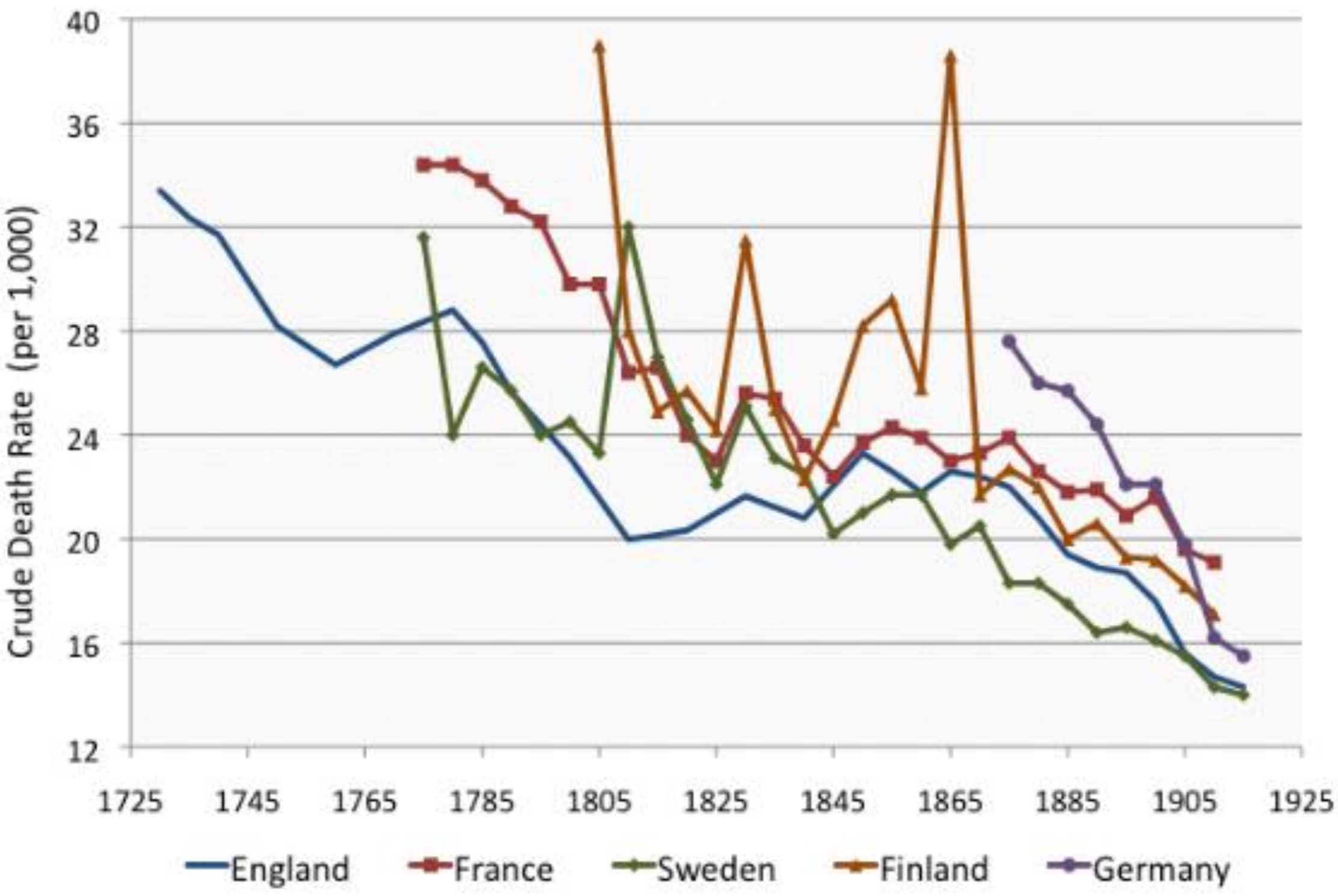

[5] The main mechanism behind the early French decline appears to have involved reduced probability of marriage per unit time, which led to more women marrying later or not at all (Wrigley 1985, p. 47).

FIGURE 6. BIRTH RATE OVER TIME IN EUROPEAN COUNTRIES, FROM GALOR (2012)

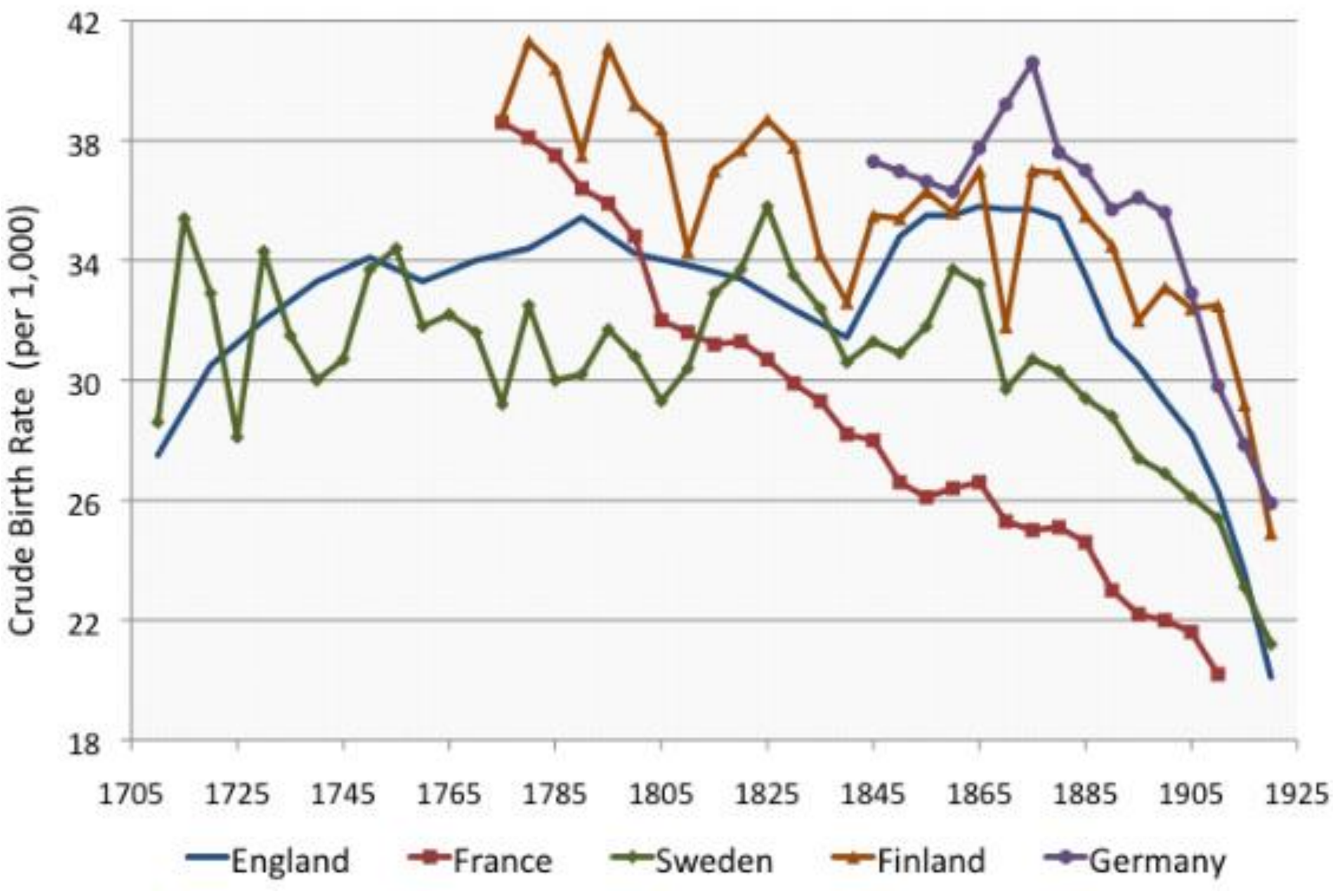

## 4 Cross-country evidence

In the graphs above, the country is the unit analysis rather than the state, region, or family. There is one number for each country, year, and variable of interest (births or deaths). One popular strategy in the study of the mortality-to-fertility link has been to bring formal statistical methods to such data, mostly focusing on the post–World War II era, for which data are better.

In fact, having come into vogue the 1990s, the cross-country approach is now somewhat in disrepute.[6] Compared to studies that track tens of thousands of families in a homogeneous district, which we will come to, studies of 50–100 developing countries have small and diverse samples. Smallness reduces statistical power, which is the ability to distinguish any patterns founds from the products of pure chance. Diversity—Ukraine differs from Bangladesh in dozens of important ways—increases the risk that omitted variables ("factors not controlled for" in the graphs above) are influencing mortality and fertility unbeknownst to the researcher. Such factors can be controlled for to the extent that they have been quantified—researchers have produced indexes of democracy, for example. But adding controls further depletes statistical power. So researchers must limit their control sets; this need increases the incentive to "mine" for the combination of controls that gives the best results, and reduces the credibility of what is published.[7]

### 4.1 Schultz (1997), "Demand for Children in Low Income Countries," *Handbook of Population and Family Economics*

A noted example of this literature is the paper already dissected, Schultz (1997). Regressions that control for several confounders and instrument with calorie intake produce the finding that for every percentage point reduction in the fraction of children who die before their fifth birthday, women had 0.25 fewer births.[8]

[6] Most of the criticism has centered on regressions in which economic growth is the variable to be explained. E.g.: "Although I applaud the empirical emphasis in recent work on economic growth, I am not sanguine about the future of this work" (Mankiw 1995, p. 307).

[7] Mining can occur in many ways, conscious and unconscious. Journals, for example, may favor papers with seemingly statistically significant results.

Taking a representative fertility rate during the study period of 4.5 births/woman (the rate was 4.6 in 1975–80 and 4.2 in 1980–85 for developing countries, UN Population Division 2013), this means that a 1% reduction in child deaths—an average 0.045 children saved per mother—caused a fertility reduction 0.25/.045 = 5.6 times as large. The estimated effect in this statistical analysis, in other words, is comparable to the 1:4 correspondence in the cross-country graph above.[9]

No one can know the exact degree to which the mortality-fertility correspondence in Schultz (1997) truly reflect causality from deaths to births. What is certain is the assumption required to believe that interpretation. As explained earlier it is: *after controlling for GDP/capita, male and female education levels, % of labor force in agriculture, and distribution among major religious denominations, a country's average per-capita calorie intake is causally linked to fertility only via mortality. In particular, no omitted variables simultaneously influence calorie intake and fertility.*

Given the general disrepute of cross-country studies, and given the size of the correspondence—saving 1 life prevents more than 5 births—I find the result hard to take at face value. That suggests other causal stories are at work, which may well be masking the one of interest.

## 4.2 Conley, McCord, and Sachs (2007), "Africa's Lagging Demographic Transition: Evidence from Exogenous Impacts of Malaria Ecology and Agricultural Technology," NBER working paper

A study appearing 10 years later, Conley, McCord, and Sachs (2007), attempts to improve on Schultz (1997) and other earlier work by instrumenting mortality something more plausibly exogenous—i.e., a factor more surely disconnected from fertility except via its power to affect mortality.

One version of this instrument is an index of malaria ecology, which represents the fraction of a country's area with biophysical characteristics such as temperature, elevation, and rainfall that favor malaria endemicity (Kiszewski et al. 2004). This version looks especially exogenous. Temperature and elevation are not affected by female education or GDP/capita.

Yet malaria ecology is not randomly distributed around the world. More precisely, its distribution is concentrated in the poorest nations and so is related to national traits that could themselves affect fertility (Kiszewski et al. 2004, p. 491, Fig 2):

[8] Schultz (1997, p. 398, Table 5, col 3). The coefficient on child mortality to age 5 is 0.0251. Since child mortality is measured per 1000 births (p. 419), a 1% reduction reduces total fertility by .0251 × 10 = .251 births/woman.

[9] Regressions in Table 6, cols 2 & 6, produce essentially the same result. A regression in Table 7, col 2, restricting to 1988 and controlling for oral contraceptive prices, produces a somewhat lower number of 0.21. Regression in cols 4 & 6 of Table 9 produce much larger estimates, but Schultz (1997, p. 413), does not find the overall pattern of results from these regressions "plausible."

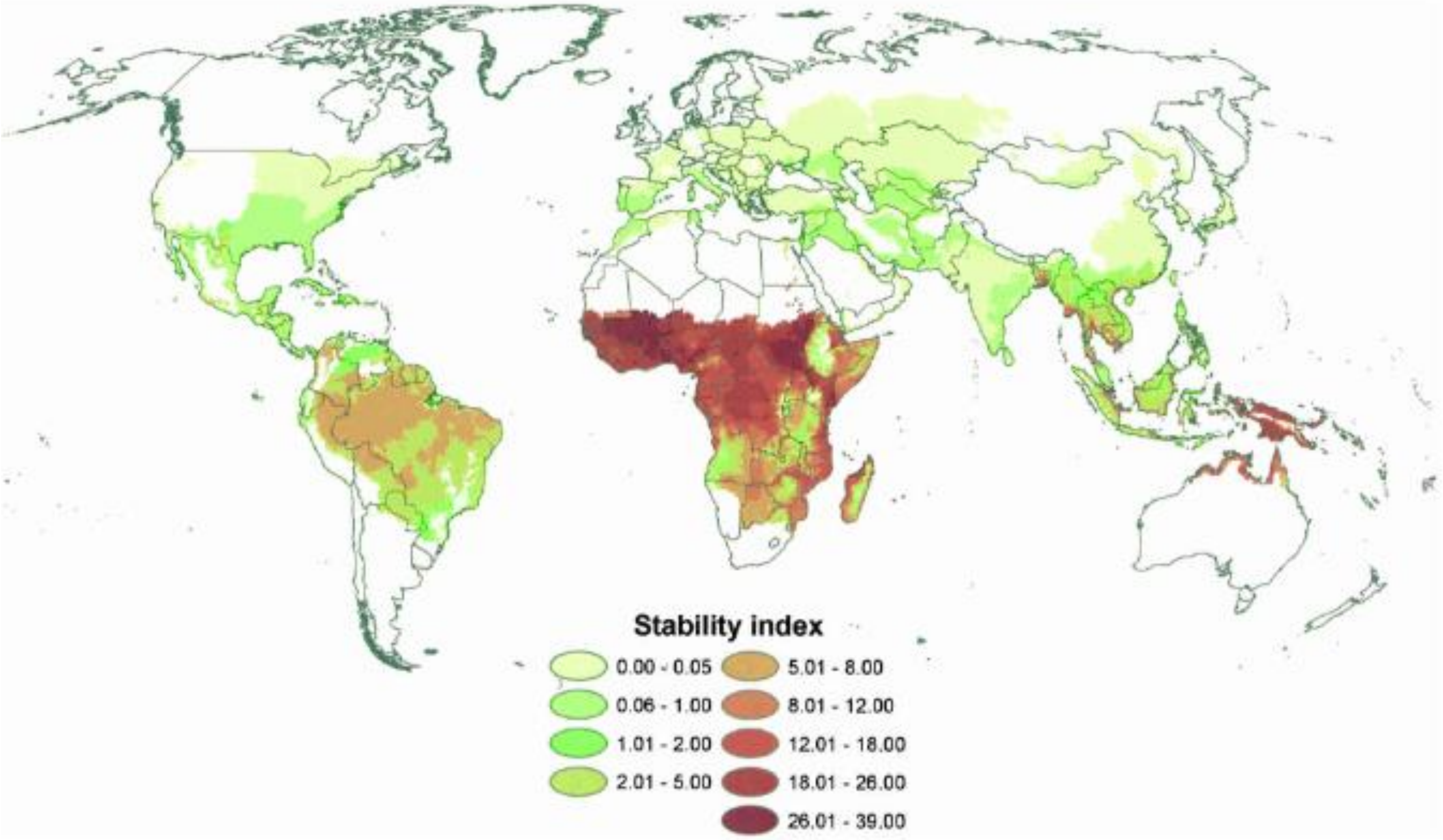

The causal map for this instrument looks like this:

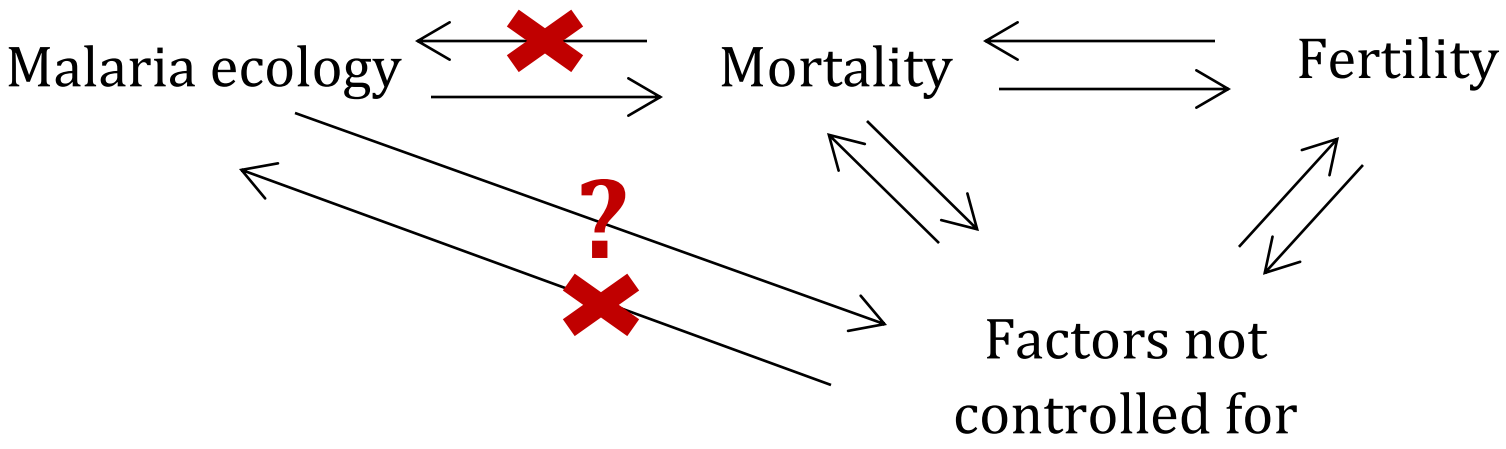

These leaves two stories to link mortality and fertility, only one of which involves the first affecting the second:

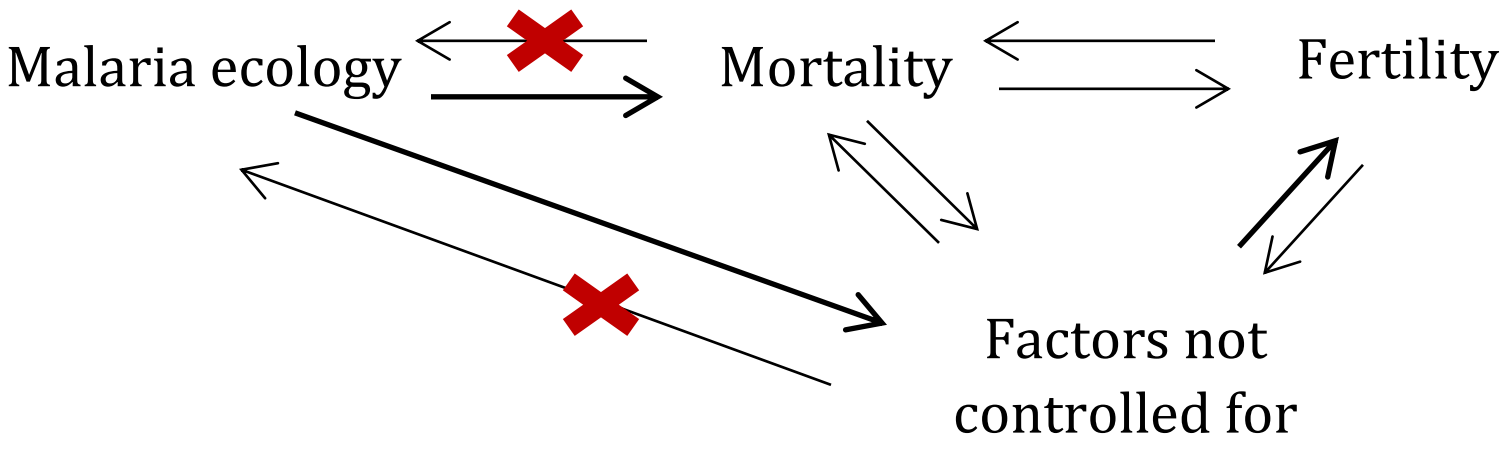

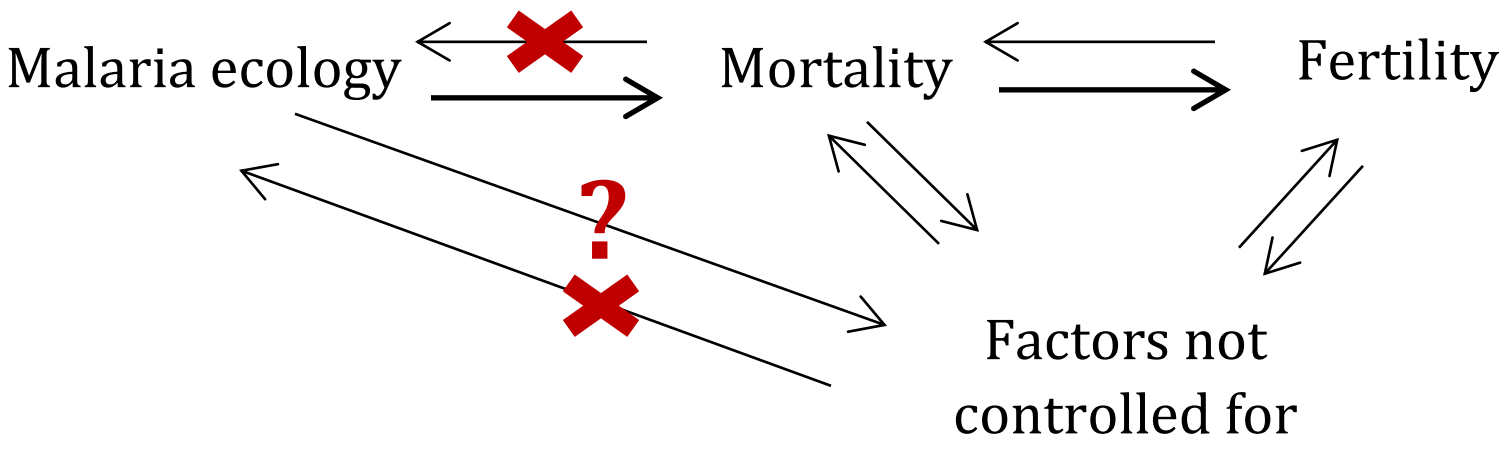

To put the upper version into words, the malaria ecology index is highest in tropical areas, where, on average over 1960–2004, fertility was also high. Even after controlling for measurable traits such as GDP/capita, it seems debatable to conclude that malaria ecology is by far the best explanation to explain elevated mortality and fertility. One can easily nominate some national trait not controlled for as invisibly correlated with malaria and influencing fertility, creating that false appearance of causal connection. The industrial revolution started in a non-tropical region, perhaps by chance, perhaps not, so tropical regions are poorer and have higher fertility. That doesn't mean, the argument would go, that malaria raises fertility.

The second version of the malaria instrument in Conley, McCord, and Sachs (2007) is a measure of actual risk of contracting malaria as a function of place and time. In additional to ecological factors, this variable depends on societal efforts such as a history of DDT spraying and swamp draining, making it less plausibly exogenous. Its advantage is that it varies over time. Statistically, this allows the authors to shift to viewing their quasi-experiment as playing out over time as well as space. E.g., instead of asking whether countries with worse malaria ecology, thus more deaths, had higher fertility during 1960–2004, they can ask whether, within countries, fertility fell as malaria risk fell. This allows them to introduce "fixed effects," i.e., to control for any trait of a country whose impact on fertility is (nearly) constant over time (see next section for more). Ironically, it also controls away the more-exogenous malaria ecology traits that are the basis of the first version of the malaria instrument.

Using the malaria ecology version of their instrument, Conley, McCord, and Sachs (2007, p. 48, Table 4, col 2), finds an impact about 40% as large as in Schultz (1997): reducing infant mortality by 1 percentage point—saving 0.0489 infants/woman since average births/woman in the study is 4.89 (Conley, McCord and Sachs, p. 41)—cuts lifetime births by 0.1 (only one significant digit is reported), for a 1:2 ratio instead of 1:5. But the impact is still large: taken at face value, the finding implies that living in malaria-prone zones *increases* population growth. For reasons already given, it is unclear whether that should be taken at face value.

The impact based on the second, less credible instrument appears to be bigger, but is not directly comparable because it is expressed per infant (under-1) rather than child (under-5) mortality, and infant mortality is always lower.[10]

## 4.3 Lorentzen, McMillan, and Wacziarg (2008), "Death and Development," *Journal of Economic Growth*

Lorentzen, McMillan, and Wacziarg (2008) includes similar cross-country regressions. They copy the malaria ecology instrument of Conley, McCord, and Sachs 2007. And it adds more instruments: indicators of geography, such as land area, distances from the equator and the nearest coast, and percentages of a country's area in various climatic zones. The paper includes a causal map (p. 97):

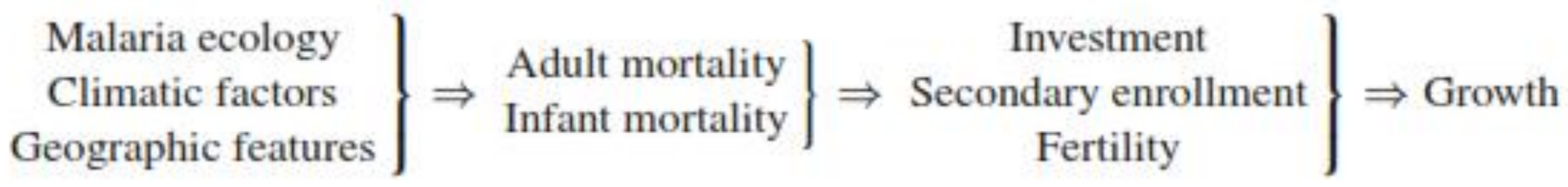

(where "Growth" refers to economic growth). As indicated, the key assumption is that the instruments affect fertility and other outcomes *only* via adult and infant mortality.

Lorentzen, McMillan, and Wacziarg (2008) finds that a percentage point reduction in infant mortality reduced births per woman by 0.15 on average during 1960–2000 (Lorentzen, McMillan, and Wacziarg

[10] Conley, McCord, and Sachs (2007, p. 50, Table 6, col 2), puts a coefficient of 0.06 on infant mortality. Infant mortality is used because of a reported lack of child mortality data for many years in the study period.

2008, p. 107, Table 9, col 6). Applying this figure to the study's average of 4.183 births/woman produces a ratio of 1% × 4.183 = .04183 to 0.15, or about 1:3. That impact, being between the results of the previous two studies, again seems large.

### 4.4 Summary: First-generation cross-country evidence

Overall, these cross-country studies produce results similar in magnitude to what one would get from the folk regression of the long-term fertility trend on the long-term mortality trend. It is unclear how well the studies have isolated the component of this broad historical relationship that is causation of fertility by mortality.

## 5 Cross-country panel studies

The three studies just reviewed share two traits:

- They do not allow for dynamic effects such as mortality affecting fertility after a lag of some years. They compare mortality in a given period to fertility in the same period. But we might expect a lag if, for instance, parents gradually notice that fewer children in their community are dying, become convinced the change is permanent, and then have fewer kids themselves. Similarly, the fertility level today might depend on the fertility level in the past, since if it was already low, it is less likely to fall much more. When left out of the analysis, past levels of fertility and mortality are potential confounders. For example, a low fertility rate five years ago might lead to a lower mortality rate today (as fewer children compete for the food supply) even as it foreshadows a lower fertility rate today (as family planning practices show some inertia). This is yet another story producing a correlation between mortality and fertility today without any impact from the former to the latter.
- The studies rely mostly or wholly on variation across countries, not over time. Staying with Lorentzen, McMillan, and Wacziarg (2008) as an example, the regressions using the malaria ecology instrument view history as an experiment in which each country is a roll of the dice. Some countries are arbitrarily endowed with conditions favoring malaria, and some not. The paper checks whether countries so endowed have lower fertility. A general fear about such regressions, as explained before, is about the failure to control for national-level confounders.

  The alternative statistical strategy has been to collect data on countries for many time periods, say, every 5 or 10 years. Then researchers can ask: within each country, when mortality falls, does fertility fall too, perhaps with a delay? This compares not Brazil to Zimbabwe but Brazil in 1990 to Brazil in 2000. It allows researchers to control for a whole class of potential confounders: those that are fixed over time or, in practice, nearly so. These include climate, culture, legal tradition, and linguistic and religious composition. More precisely, it allows researchers to control for these factors to the extent that their impact is fixed over time—what are called "fixed effects." For if the impact on fertility of having many Catholics is fixed, then that impact washes out when examining changes over time within a country.

One thrust in modern econometrics has been to depart from those two common characteristics, to allow for dynamics and fixed effects. Both steps sharpen the focus on how variables interrelate over time. They require more complicated econometric methods; they pre-process the data to remove fixed effects and/or post-process the regression results to infer the long-term evolution arising from dynamic interactions among variables. The methods are called "panel" methods because they use data collected at regular intervals from a set of families, firms, or countries—in our case, a "panel" of countries followed over time.

Two main approaches have developed in the last 20–30 years: dynamic panel data models; and vector autoregressions. I will review one leading, recent example of each, explaining the methods in brief.

## 5.1 Murtin (2013), "Long-term Determinants of the Demographic Transition," *Review of Economics and Statistics*

Murtin (2013) works with decadal data series from 70 countries, some reaching back to 1870. The variables of primary interest to us are the national birth and death rate, as in the Galor (2012) graphs above, and the infant mortality rate.[11] Controls include GDP/capita; the average number of years of primary schooling that adults have completed; and the fractions of the population in their 20s and 30s, respectively, since a younger population will have more children and more child deaths.[12] Most of the fixed-effect regressions—the ones controlling for national traits whose impact on the birth rate stays the same over time—find that a 1% reduction in infant mortality (not 1 percentage point) is associated with a 0.3% (not 0.3 percentage point) drop in the birth rate (Murtin 2013, p. 624–25, Tables 4–6).

Interpreted as causation from mortality to fertility, this statistic again points to a large effect, which has saving lives leading to slower population growth. We can see this by taking values for infant mortality and total fertility from the center of the graph at the beginning of this document (circa 1980) and assuming that the proportional effect on the birth rate (births per population, which is what Murtin studies) is the same as on total fertility (lifetime births/woman). The figures are then that a woman averages 4.5 lifetime births, of which 9% (0.4) result in death before age 1. According to the Murtin (2013) result, reducing the 0.4 infants lost by 1%, or 0.004 children, would cause a reduction in the woman's lifetime births of 0.3%, or 0.3%×4.5=0.0135 children, for a mortality-change-fertility-change ratio of about 1:3, comparable to the earlier studies.

However, these regressions do not instrument infant mortality, so they do not try to rule out reverse causation, nor causation by (time-varying) third variables.

Murtin's dynamic panel regressions introduce two major changes and one minor one. First, being "dynamic," they control for the previous decade's birth rate ("lagged birth rate") even as they study predictors of the current decade's rate. Since fertility trends are stable over time—fertility does not jump up and down from year to year—the lagged birth rate is a strong predictor of the current birth rate. Bringing it in as a control creates a powerful competitor for mortality as explanator of the current period's birth rate. Mortality only wins to the extent that it is more correlated with the current birth rate. This makes the regressions more conservative in ascribing causality to mortality. The model is called "dynamic" because the birth rate in the present is assumed to depend on the birth rate in the past.

The flip side of this conservatism is a fundamental change in the nature of the interrelationships modeled. Imagine that the invention of antibiotics reduces mortality starting in the 1940s: it is a one-time but *permanent* drop. Murtin's dynamic model, which is standard, allows the mortality drop in the 1940s to reduce fertility in the 1940s; it then allows this change to ripple through the generations as, say, parents influence through their example the fertility choices of their grown children. One decade's fertility directly affects that of the next. The ripples do decay—after all, the fertility choices of our 18th-century ancestors don't affect us much now. But *in addition*, since the mortality drop is permanent, from a mathematical point of view, antibiotics separately cut mortality in the 1950s. According to the model, this too sends a ripple through the generations, just like the first but starting one decade later. Yet another wave starts in the 1960s, piling on top of the earlier ones. And so on. The total effect is not infinite because all the ripples are decaying as they age. In the long-term, the effect of a permanent mortality drop converges smoothly to a limit (assuming antibiotics don't become ineffective). This long-term effect can be much larger that the short-term (one-decade) effect.[13]

[11] The birth and death rates are "crude" rates expressed per population. This makes them dependent on the distribution of the population across age groups. Populations with more young adults will have higher crude birth rates, even if the populations have the same long-term reproductive rate. This is why studies on modern data usually use sharper measures of fertility (births/woman on current birth probabilities at each age and mortality (death rates for specific age groups). Studies on historical data typically do not have this luxury, for lack of requisite data.

[12] All variables except population fractions and schooling are entered in logs. Log GDP/capita enters in cubic form. Decade dummies are also included.

The second major change is that the regressions instrument most of the variables. This they do in a few different ways, which appear rather arbitrary, are never justified, and ultimately do not convince. The most credible variants for us (Murtin 2013, p. 626, Table 7, cols V & VI) assume a world essentially like this:

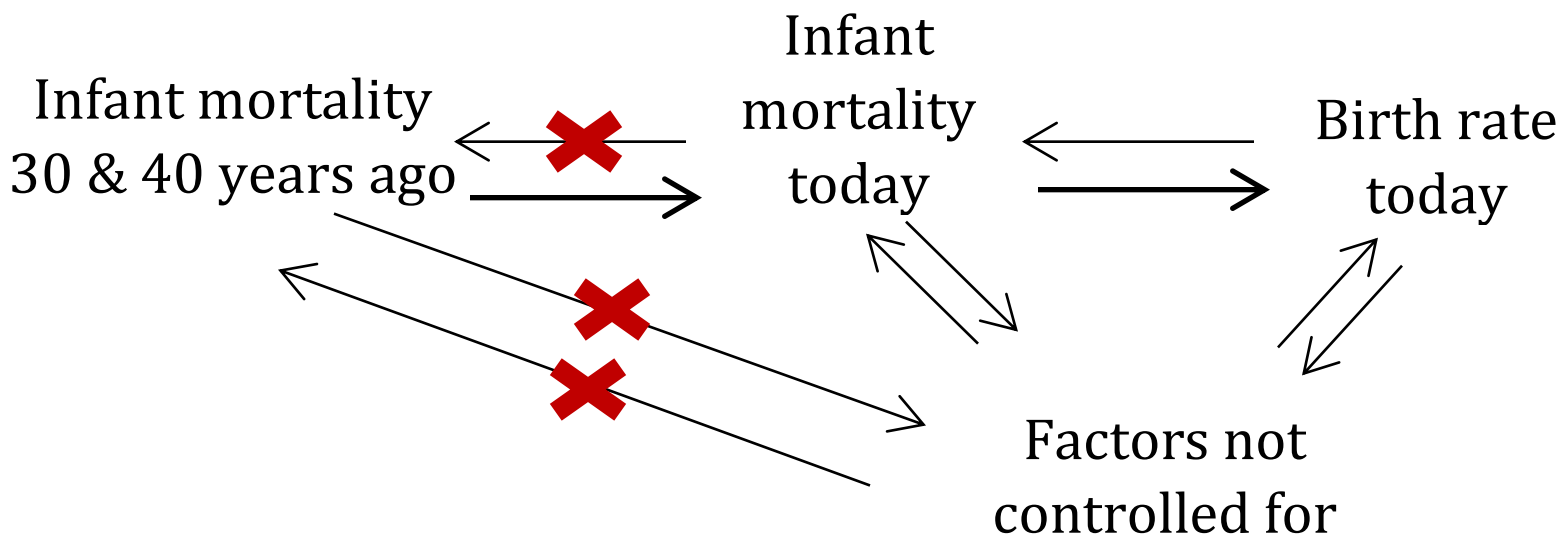

Infant mortality a generation or so ago—or the set of deeper factors it proxies, such as the state of medical practice in the country at the time—is assumed to affect current infant mortality, and *only thereby* the current birth rate. This practice of instrumenting variables with older observations of themselves is now common. It is easier to think of than a clever instrument like malaria ecology; and freely available software—in this case, a program I wrote (Roodman 2009b)—automates the implementation. The strategy is motivated by the top red X in the diagram above, the idea that current realizations of a variable can't affect past ones. But it doesn't justify the other red X's—a point that is broadly relevant in panel econometrics and broadly underappreciated. There are still ways to explain any correlations found among the variables along the top without mortality affecting fertility:

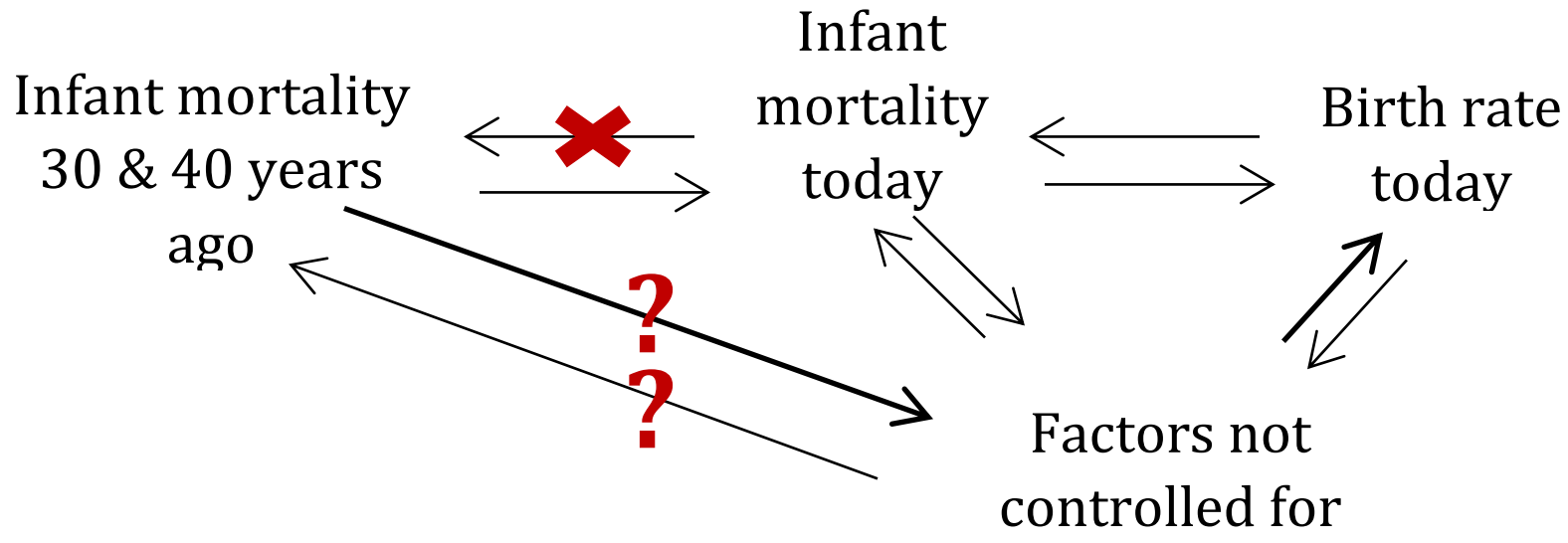

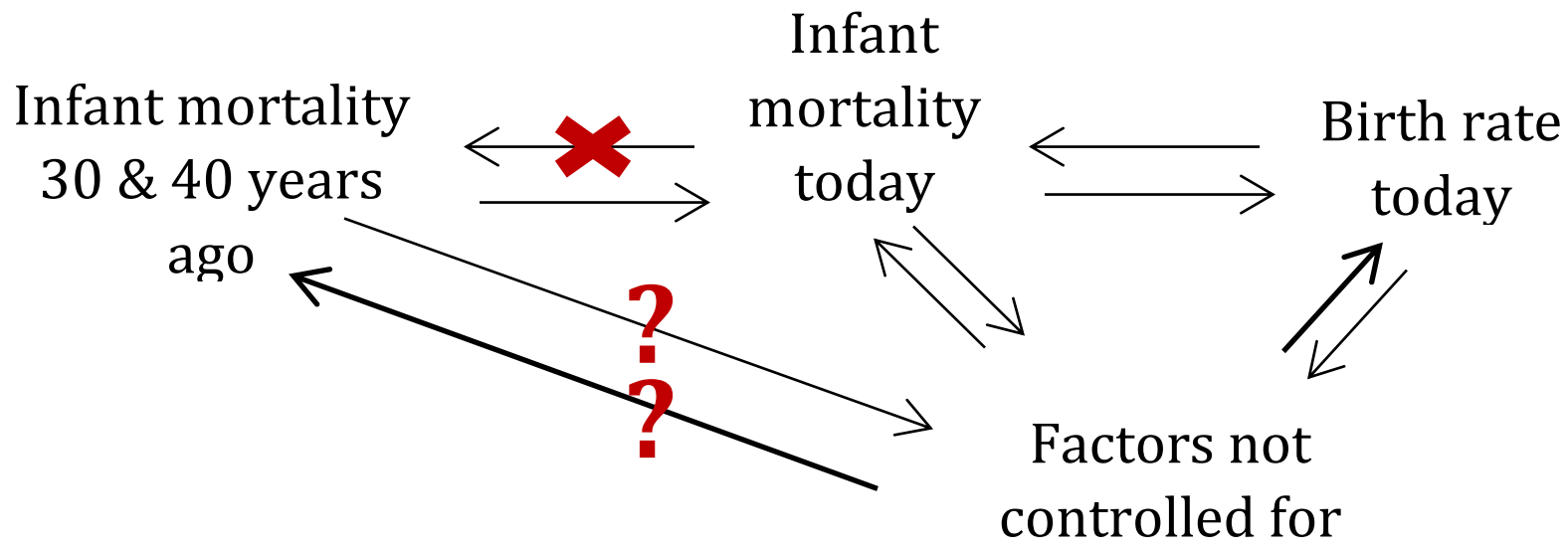

The third change that Murtin (2013) makes in moving beyond the preliminary fixed effect regressions is the least important but has the most impressive name: adopting the Generalized Method of Moments (GMM). Notice that the regressions depicted above introduce two instruments for infant mortality: infant mortality 30 years ago and infant mortality 40 years ago. Under Murtin (2013)'s assumptions, either would suffice alone. But a general principle in econometrics is that the more of the available information that is exploited, the more precise will be the results. This is essentially why Murtin (2013) uses both instruments. The use of more instruments than is strictly needed raises a mathematical question: how much weight should each get? The estimate of the impact of mortality on fertility will vary depending on the weighting, though, one hopes, not by much. GMM is a method for choosing the weights that will maximize the precision of the results, at least as the sample size goes to infinity.[14]

[13] In a dynamic model $y_t = \alpha y_{t-1} + \beta x_{t-1}$, the long-term impact of a permanent increase in $x$ on $y$ works out to $\beta/(1-\alpha)$.

Probably because of the first major change embodied in the dynamic panel GMM regressions—controlling for the previous decade's birth rate—the contemporaneous correlation between infant mortality and the birth rate drops by two-thirds in these more complex Murtin (2013) regressions. Now, a 1% reduction in infant mortality is associated with a 0.1% drop in the birth rate, instead of 0.3%. For the representative woman with 4.5 births, this cuts the mortality-change-fertility-change ratio from 1:3 to 1:1 (Murtin 2013, Pg 626, Table 7, cols V & VI). But if the 1% infant mortality drop is permanent, its long-term effect grows large again: to a 0.2--0.4% fertility drop, or a ratio of 1:2 to 1:4. The dynamic panel GMM regressions, however, appear problematic to me in several technical respects. I think the best interpretation is that the ratios just inferred represent real correspondences but that, just as in the studies previously reviewed, causality from mortality to fertility may not be the only mechanism at work.[15]

## 5.2 Herzer, Strulik, and Vollmer (2012), "The Long-run Determinants of Fertility: One Century of Demographic Change 1900–1999," *Journal of Economic Growth*

In the context of this review, the most significant and credible innovation in Murtin (2013) is the addition of the lagged birth rate as a control when studying correlates of the current birth rate. Another stream in econometrics goes much farther in that direction, as manifest in Herzer, Strulik, and Vollmer (2012).

The source of that stream is a paper by Clive Granger (Granger 1969) that confronts the question of what it means, as a matter of statistics, for one variable to cause another. He offers one definition: $x$ causes $y$ if forecasts of $y$ based on only on knowledge of past values of $y$ can be improved with information about the history of $x$. Bond investors, for example, use their knowledge of the history of interest rates to predict future interest rates. But they also use other information—say, the fact that the country has just plunged into deep recession—to improve their predictions. We would then say that recessions "Granger-cause"

[14] Roodman (2009b, p. 88–94), formally presents linear GMM.

[15] The regressions appear to suffer from instrument proliferation, manifest by Hansen overidentification $J$ test $p$ values near 1. Murtin (2013) cites Roodman (2009c) for the rule of thumb that the number of instruments should not exceed the number of countries. However, that text is much more cautionary: "The…results just cited and replications below suggest that keeping the instrument count below $N$ does not safeguard the $J$ test." Separately, Murtin's unexplained choice to instrument all variables other than time dummies with lags of just one or two probably results in weak instrumentation. Roodman (2009b) describes typical use of the xtabond2 program: "most regressors appear twice in a command line, once before the comma for inclusion in $\mathbf{X}$ and once after as a source of IV- or GMM-style instruments." I modified the publicly posted code for the two Murtin (2013) regressions of focus in the text by generating lagged instruments from all the regressors other than time dummies, and "collapsing" them to reduce their number. These regressions still put coefficients of about 0.1 on log infant mortality, with statistical significance around $p = 0.1$. The commands are: "xtabond2 lFert l.lFert fmsoto lInfantMort lDeath lyact lyactsq lyactcub p30joint p40joint d1* if OKfert==1, gmm(fmsoto lFert lInfantMort lyact lyactsq lyactcub p30joint p40joint, lag(3 4) collapse) iv(d1*) two robust" and "xtabond2 lFert l.lFert fmsotop fmsotosh lInfantMort lDeath lyact lyactsq lyactcub p30joint p40joint d1* if OKfert==1, gmm(fmsotop fmsotosh lFert lInfantMort lyact lyactsq lyactcub p30joint p40joint, lag(3 4) collapse) iv(d1*) two robust". Borderline results on the Hansen overidentification test ($p = 0.07$ and 0.15) undercut the validity of the instruments, thus the causal interpretation. Using just 4th lags as instruments improves the Hansen test somewhat ($p = 0.22$ and 0.26) but destroys the statistical significance of the results.

falling interest rates. If on average event A precedes event B than A Granger-causes B.

In many cases, Granger causality aligns with everyday notions of causality. An investment bank collapse today influences and predicts the stock market drop tomorrow. The death of a child leads a couple to have another. Sometimes, however, causality between observed variables goes backwards in time. The stock market might rise in anticipation of an interest rate cut. Cars slow down at traffic lights before they turn red. But studying Granger causality brings the great virtues of transparency and humility. If researchers report that traffic lights turn red after cars slow down, their conclusion is indisputable and should be understood to mean no more or less than its literal statement. At the same time, as the traffic light example suggests, analysis in the Granger tradition does not eliminate the fundamental difficulty of inferring true causality from non-experimental data.

Granger analysis shifts the analysis from how a large set of variables—female education, GDP/capita, etc.—affect an outcome of interest to how a small set of variables affect each other over time. A Granger analysis might look at how unemployment and short-term interest rates affect each other, and with what lags. A key uncertainty for the U.S. Federal Reserve, for example, is the time profile of the impact of an interest rate change on the unemployment rate. Perhaps the impact begins to show up after 6 months and reaches peak size after 12. Meanwhile, a key uncertainty for investors runs the other way: how will an unemployment drop play out in Fed policy? To trace the time profiles of these effects, typical Granger regressions include many lags of the variables of interest. The current month's unemployment rate, for example, is simultaneously checked for correlation with the short-term interest rate a month ago, the rate two months ago, the rate three months ago, and so on.

The results of such regressions depict causal spirals: unemployment in one year affects interest rates in the next, which perturbs unemployment the year after that. Often, researchers are interested in how an unexpected or unprecedented change in one of the variables—an investment bank meltdown, a DDT spraying campaign—will ricochet through such a system of variables that affect each other with time delays. Standard methods and software provide the answer by crunching the regression results and plotting "impulse response functions," graphs that show how one variable deviates from its otherwise expected path in the months and years after a sudden change in itself or another variable.

Herzer, Strulik, and Vollmer (2012) is a Granger-style study based on data from 20 countries for 1900–99, with observations taken every 5 years. Most of the countries are today classed as high-income by the World Bank—Canada, Japan, Western European states, Chile, and Uruguay. Three are considered upper-middle income: Argentina, Colombia, and Venezuela. Only Sri Lanka is considered poorer (World Bank 2014).

The regressions in Herzer, Strulik, and Vollmer (2012) that are key for our inquiry examine the correlations among just three variables: the birth rate and death rate per thousand of population, and GDP/capita. Each variable is allowed in the model to be influenced by the values of all three in the previous two periods: the average values in the previous 5 years, and in the 5 before them (Herzer, Strulik, and Vollmer 2012, p. 371).[16] Since there are three variables potentially influenced by the past, three variables to exercise the influence (the same three), and two periods over which influence may be detected, these regressions produce $3 \times 3 \times 2 = 18$ numbers. Examples: the correlation between the birth rate 5 years ago and the birth rate now, the correlation between mortality 10 years ago and the birth rate now; the

[16] GDP/capita is taken in logarithms. The regressions focused on here also include an "error correction" term, which is the deviation of the birth rate in the previous period from the level predicted by a separate regression determining the overall, long-term correspondence between the three variables. This deviation represents innovations in fertility—changes not predicted by past levels of any of the variables—and it proves a statically significant correlate of future values of all three variables, meaning that it Granger-causes mortality and GDP/capita. Herzer, Strulik, and Vollmer (2012, note 16), states that specifications with 1 or 3 lags instead of 2 produce qualitatively similar results.

correlation between GDP/capita change 10 years ago and mortality change now.[17]

Herzer, Strulik, and Vollmer (2012, p. 373), do not report those numbers, but they graph what happens when the numbers are fed into a simulation of the dynamics. This is how the birth rate evolves after an unexpected jump in the death rate:

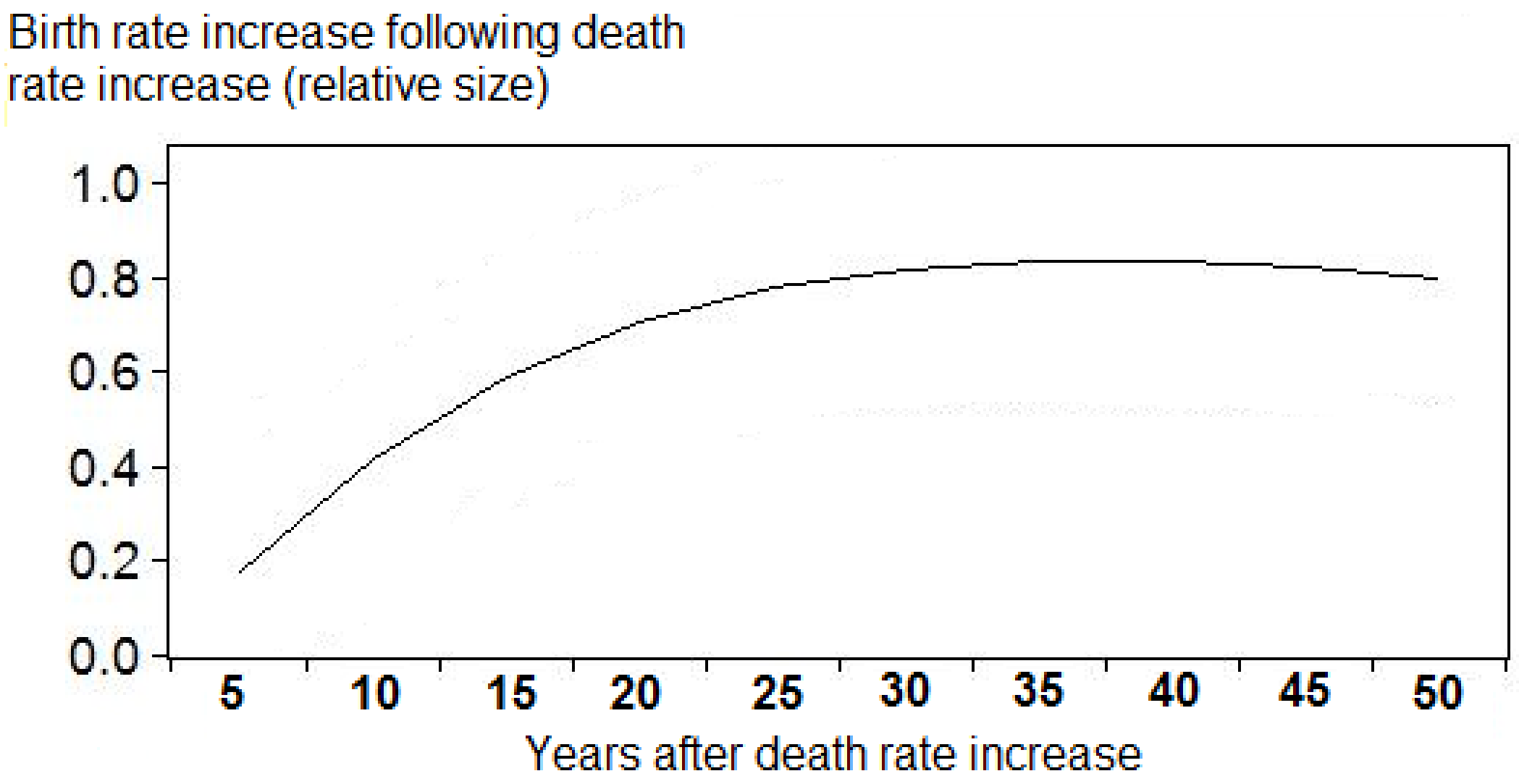

Going by the patterns in this data set, a one-time jump in the death rates led over about a generation to an increase in the birth rate. The full Granger effect peaks at 80% of initial jump in the death rate. Thus, a reduction in the death rate is eventually followed by a fertility reduction 80% as large, meaning a permanent but modest increase in population growth. The 1:0.8 mortality-change-fertility-change ratio from this rich dynamic model is not much lower than the 1:1 ratio from Murtin's simple one, and far lower than the non-dynamic studies in the previous section.

To repeat, studies in the Granger tradition tell us what follows what—not what causes what, in the everyday sense of "cause." This one tells us that in the 20th century, in these mostly-wealthy nations, drops in mortality that would not have been expected based on historical trends have been followed, after a generation, by slightly smaller drops in fertility.

### 5.3 Summary: cross-country panel studies

Though emerging from different econometric traditions, these two panel studies share some important traits. Both focus on changes over time within countries and allow for dynamics. Of the two, Murtin (2013) emphasizes using instruments to pin down cause and effect, and does not completely succeed. Herzer, Strulik, and Vollmer (2012) gives up on that task and instead more fully models the dynamic relationships among variables. It estimates the correlation between mortality drops and subsequent, long-term fertility changes at about 1:0.8—notably smaller than the other studies so far scrutinized.

## 6 Quasi-experimental studies

Recognizing the difficulty of pinpointing cause and effect through instrumental variables studies like those reviewed above, economists have turned in the last 15 years to performing experiments, as well as searching for and exploiting "quasi-experiments" such as the sudden introduction of antibiotics.

I am aware of no experimental studies linking mortality to fertility. This is perhaps unsurprising given the ethical problems in randomly depriving some people of a potentially lifesaving intervention.[18] But several studies have appeared that attempt to take advantage of quasi-experiments affecting mortality in order to quantify the knock-on effects for fertility.

[17] By "correlation" here I mean partial correlation.

It should be said that "quasi-experiment" is only quasi-defined. Arguably any instrumental variables study is a quasi-experiment. Schultz (1997), for example, takes advantage of allegedly arbitrary cross-country differences in the instrument, per-capita calorie intake. What distinguishes the new crop of studies is a reliance on external developments such as the HIV pandemic and the global push for child immunization, which trigger sudden changes in the health regime within states, regions, or countries. Like the panel studies described earlier, these quasi-experimental studies focus on changes over time rather than purely differences across space. And because of their focus on external events, their claims to quasi-experimental status are more credible. Still, just like other studies, quasi-experimental ones make assumptions—exclusion restrictions—that are not beyond debate.

## 6.1 The convergence problem

Most of the quasi-experimental studies reviewed here focus on episodes in which a sudden improvement in the health regime—DDT spraying for malaria, the arrival of antibiotics or vaccines—occurs over a large area. In the years that follow such events, different geographic regions realize different degrees of decline in their death rates. Researchers analyze whether where deaths fell, births fell too.

These episodes look "quasi-experimental" in two ways. First, their timing is usually not a function of local events—in particular, not a function of the local fertility rate or third factors such as inequality and GDP/capita. DDT spraying, for example, went global in the 1950s because of a combination of scientific advance and advocacy by international players [cite]. Second, the *suddenness* can create useful statistical discontinuities. Sri Lanka in 1945 was probably statistically similar to Sri Lanka in 1950 *except* that a national malaria eradication had been launched in between. A before-after comparison tightly bracketing a major health regime change is more experiment-like than a comparison over 1 longer period. It is harder to argue that a third factor such as falling inequality obscured the true mortality-fertility link if that third factor could hardly have changed in such a short period.

Studies exploiting such episodes do suffer one weakness: the episodes usually cause convergence. At the end, malaria or antibiotic-treatable diseases are gone where they were once prevalent *and* where they never were. Thus the mortality drops can be statistically indistinguishable from the initial mortality levels, which are not random.

A good illustration of the conceptual issue (though not so relevant to this review because it does not cover fertility) is the Bleakley (2010) assessment of the economic impacts of anti-malaria campaigns in Brazil, Colombia, Mexico, and the United States in the 20th century. Consider the case of Brazil. The map on the left of Figure 7 shows the Kiszewski et al. (2004, p. 491) malaria ecology index—the same one used in Conley, McCord, and Sachs (2007) and Lorentzen, McMillan, and Wacziarg (2008)—for Brazil. A higher number (toward brown) indicates a zone more favorable to malaria. Brazil's malaria ecology follows a strong pattern, with inland areas more prone than coastal ones, which are themselves more urbanized and industrialized and less poor, as the map on the left shows (Ferreira Filho and Horridge 2006). Those same inland areas are probably marked by different histories, cultures, and economic circumstances.

[18] Nevertheless, such an experiment is not inconceivable. Cohen and Dupas (2010) randomize the price of bed nets, which could affect mortality and fertility, although they do not track these outcomes.

**FIGURE 7. MALARIA ECOLOGY AND POVERTY RATE ACROSS BRAZIL**

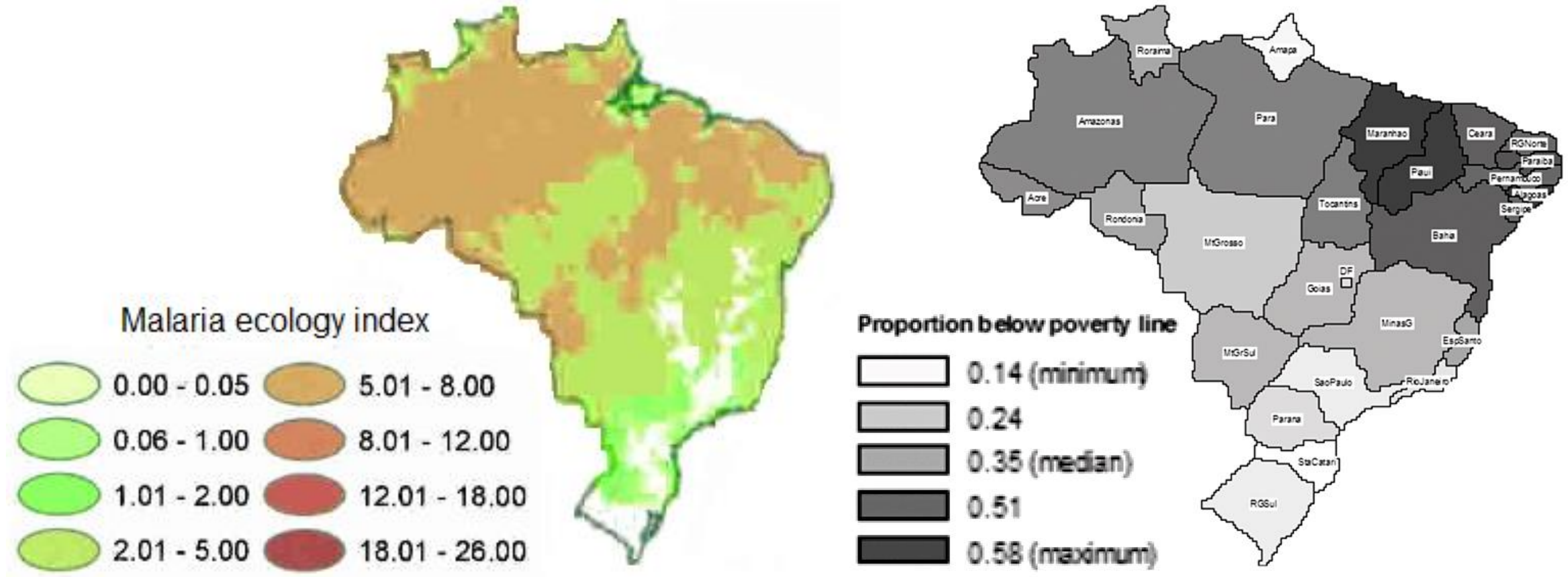

Bleakley (2010) uses this index for Brazil, in studying the impacts of a mid-century malaria eradication campaign built around DDT. The geographical pattern enters the left end of the Bleakley (2010) causal graph this way, with the unit of analysis being the Brazilian state:

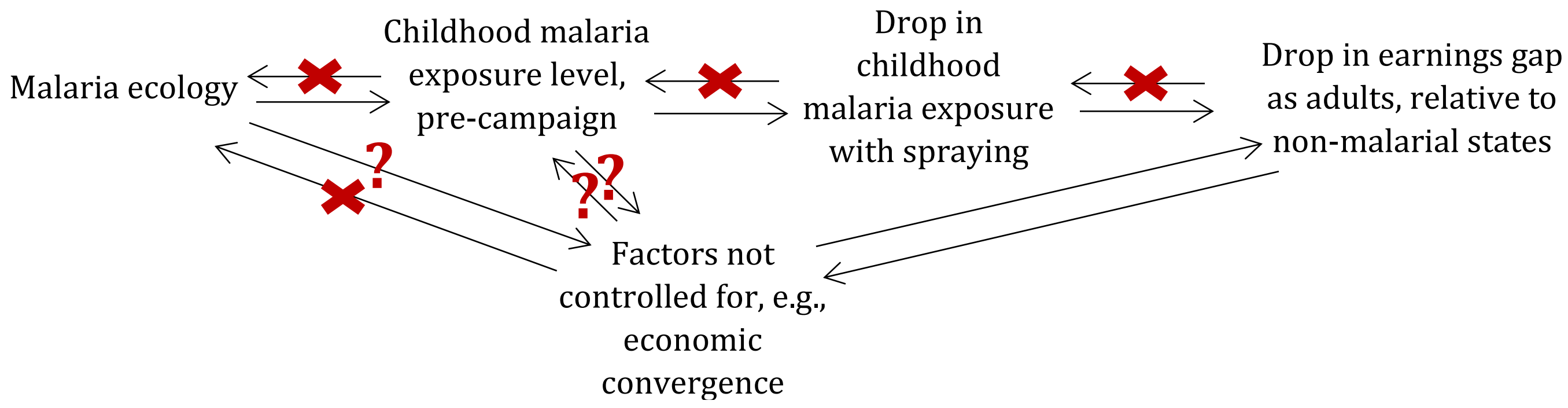

Bleakley (2010) favors the causal story running along the top: geography influences pre-campaign malaria levels among children, which influences subsequent drops, which influences changes in the earnings gap between natives of malarial and non-malarial regions. But as shown, other causal pathways could link geography to a narrowing of the earnings gap. One candidate for a hidden factor not controlled for would be a long-term pattern of economic convergence among regions. Maybe the poorer, more malarial states, just caught up economically over the decades of the mid-20th century for reasons separate from malaria eradication.

However, Bleakley does something clever to attack this competing theory. He collects follow-up data at many points in time (yearly) after the intervention on his outcome of interest, which is earnings as adults of children who grew up in states with different malaria ecology levels. This is useful because while economic convergence and DDT spraying might both have narrowed the earnings gap, the latter would probably have done so with distinctive timing. The narrowing should commence just as the first children whose malaria exposure was reduced by the spraying campaign enter the workforce—perhaps they were 15 when the spraying campaign started—and continue steadily until all entering the workforce grew up minimally exposed. The data seem to confirm this pattern. This graph shows the gap in future earnings between people from malarial and non-malarial regions by year of birth (Bleakley 2010, p. 26). The figures start negative, indicating lower earnings for those from malarial regions:

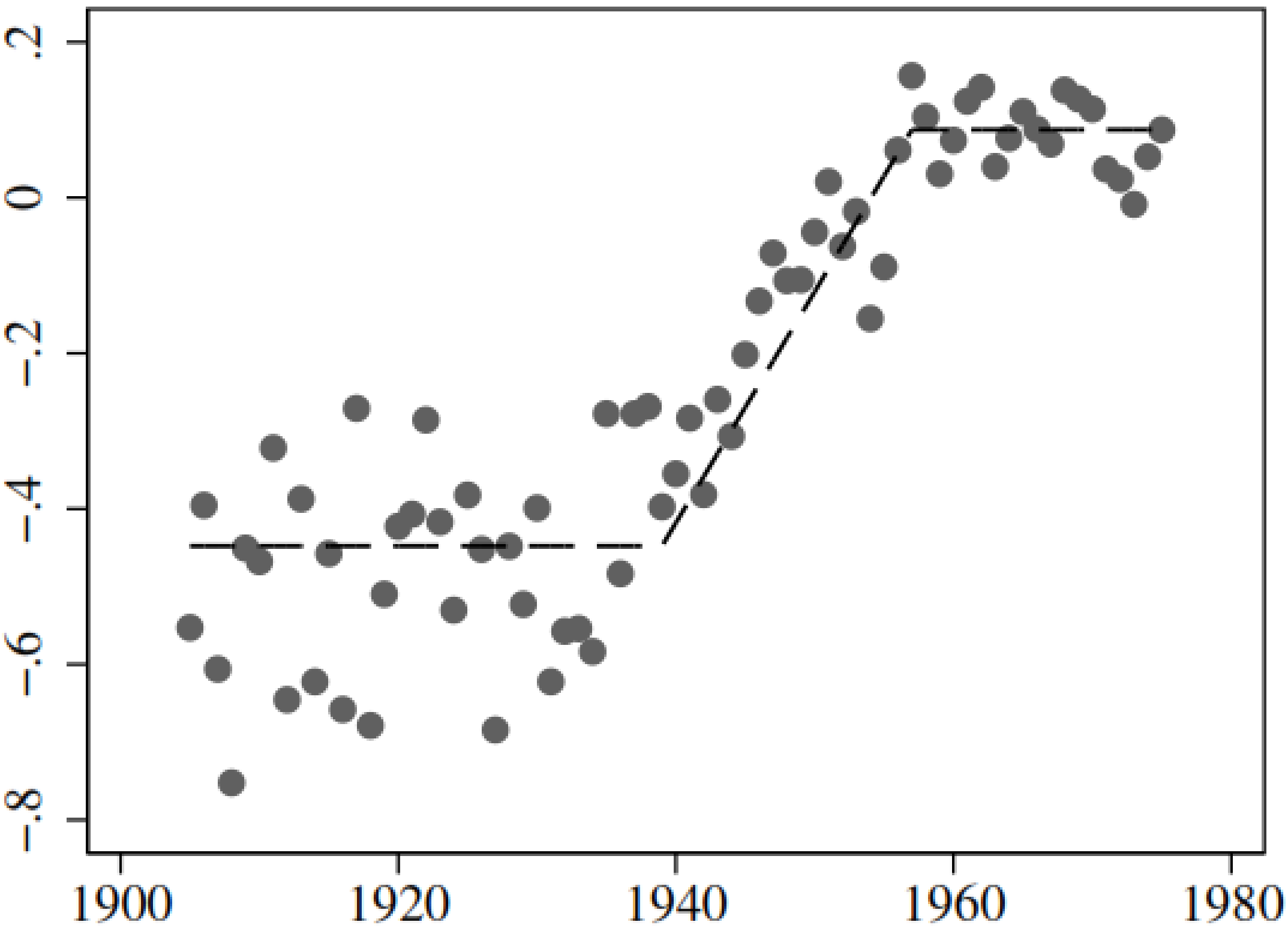

Children born in malarial areas before 1940 reached adulthood without ever benefiting from the national DDT spraying campaign that began in the mid-1950s. Their future earnings were systematically lower according to the left part of the graph. This finding is consistent with the theory that malaria impedes children's cognitive development, for reasons of biology and lower school attendance, and thus reduces their adult earning power. Meanwhile, children born after 1960 in ecologically malaria-prone fully benefited from eradication, and for them the earnings gap disappeared, even reversed slightly.

One might argue that the kinks are an optical illusion and that a straight line fits the data nearly as well. But Bleakley (2010, p. 23, Table 4, Panel B), reports statistical tests that suggest otherwise. The point for this review is that Bleakley's graph is more much more convincing than it would be if it had only two data points, from before and after eradication. For then there can be no distinctive fingerprint:

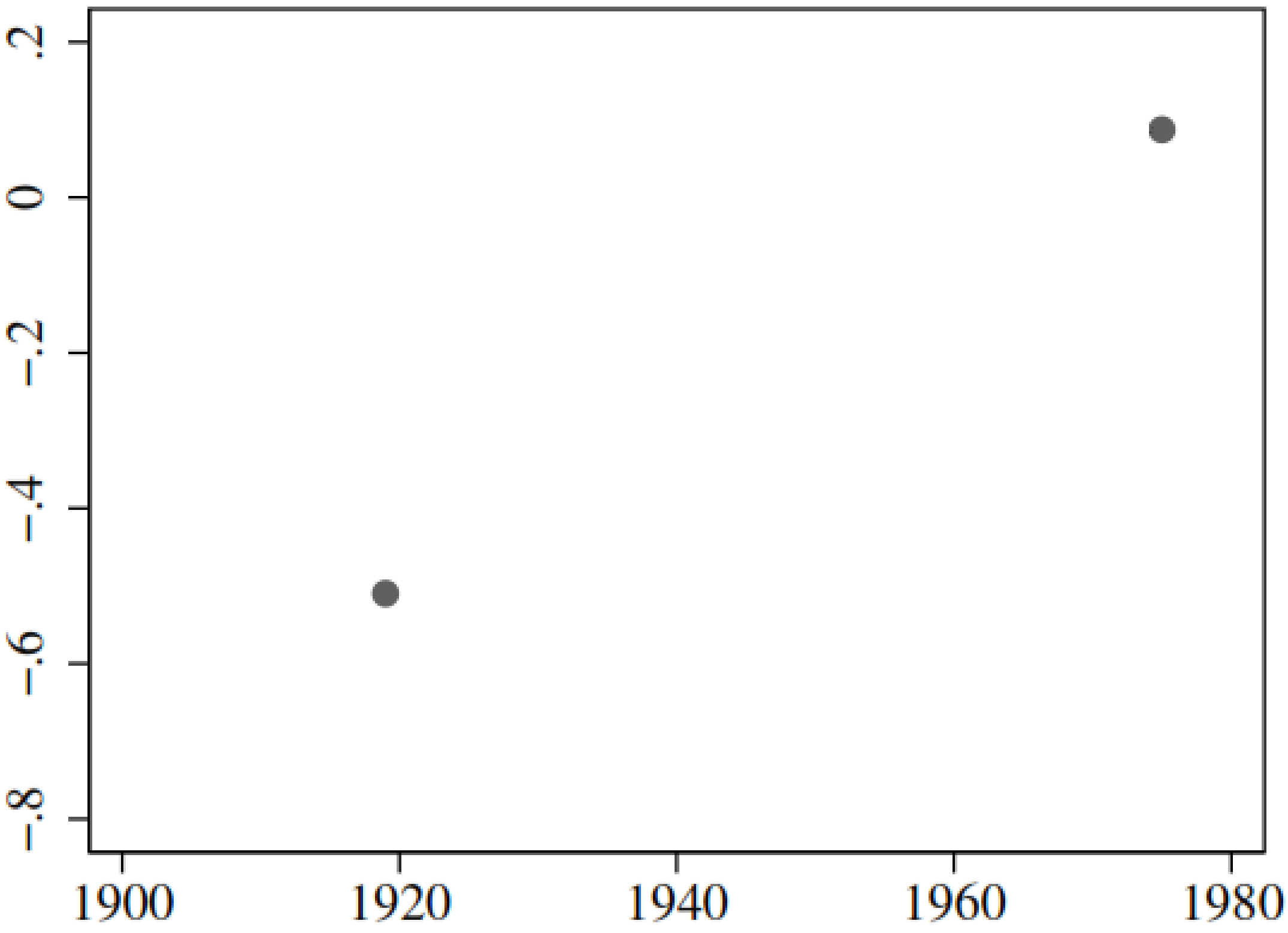

Quasi-experimental studies that look for fingerprints in time series, rather than just performing before-after comparisons, are more convincing.

## 6.2 Lucas (2013), "The Impact of Malaria Eradication on Fertility," *Economic Development and Cultural Change*

Among the quasi-experimental studies reviewed here, Lucas (2013) comes closest to the ideal implied above of checking for a temporal signature. Rather like Bleakley (2010), Lucas (2013) studies a malaria eradication episode—this one in Sri Lanka circa 1947. The outcome data come from a survey of 6,810 women conducted in 1975, which gathered detailed birth and child death histories from each respondent. Many of those births occurred before the spraying and many after. This data structure allows Lucas (2013) to examine the evolution of birth and death probabilities year by year, in order to discriminate between general trends and developments more precisely correlated with eradication.

Lucas (2013)'s measure of how malarial a region was before eradication is the "spleen rate," that is, the fraction of children with an enlarged spleen, which indicates a history of exposure to malaria. In the late 1930s the spleen rate varied by region between 1.5% and 68% (Lucas 2013, p. 611). By 1960, it was essentially zero throughout Sri Lanka (Lucas 2013, p. 612). The graph below (Lucas 2013, p. 619) shows how the probability of a woman having a baby in a given year evolved in the once-malarial areas. A value on the graph of, say, 0.2 in 1950 does not mean 20% of women gave birth in 1950. Rather, it means that if a woman lived in an area that in 1937 had a spleen rate of 50%, her chance of giving birth was in 1950 0.2 × 50% = 10% higher than it had been back in 1937. It represents, in other words, the impact of the passage of time on fertility in malarial areas relative to non-malarial areas. The solid line in the graph traces the best estimates of this impact year by year. Since such statistics are always measured with uncertainty, the dashed lines indicate confidence intervals, ranges within which Lucas (2013) is 95% sure the true value lies:

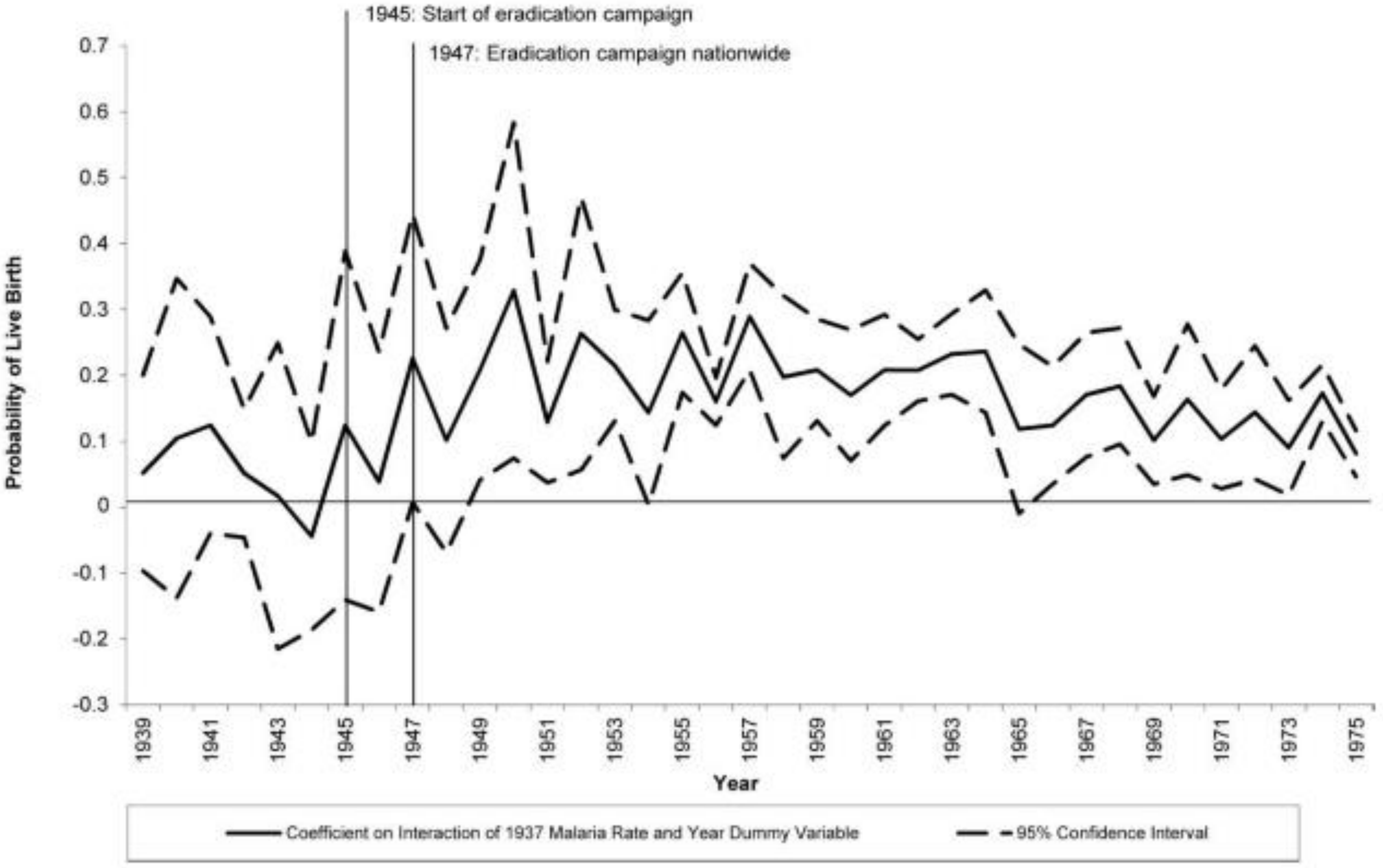

Notice that before the nationwide eradication begins in 1947, zero lies well within the 95% confidence intervals: there is no clear change in relative fertility in 1937–47. But soon after, the difference rises to statistical significance, as the confidence intervals then exclude zero. The difference appears fairly stable around 0.2 through the mid-1960s. Its slight decline thereafter might reflect a general compression in differences across Sri Lanka as fertility fell overall. This looks like a fingerprint of malaria eradication.

Lucas (2013) backs the observations above with formal regressions. These control for district fixed effects, individual-level traits such as ethnicity and, most importantly, district-specific linear (straight-line) trends (Lucas 2013, p. 613, Eq 1). As in the Bleakley (2010) Brazil analysis, the regressions check whether allowing a particular nonlinear pattern, a jump in 1947, can explain the data better than straight-line fertility trends alone. Consistent with the graph, the probability of a live birth is 0.22 higher after 1937 (Lucas 2013, p. 620, Table 2, col 1).

Notice that the rise in fertility is the opposite of what we would expect if thought that the primary causal story was from eradication to lower mortality to the volitional channel of parents choosing to have fewer children. But it is compatible with a direct biological link from malaria to fertility. Malaria can cause spontaneous abortions and stillbirths, which translate statistically into lower fertility. This effect may be distinguishable in the data because it mostly tells on a woman's first pregnancy (Lucas 2013, p. 608). When Lucas (2013) splits the data between first births and later births, the clearest impact of malaria eradication is indeed on first births. The estimates of the impact on fertility thereafter are unstable and overall indistinguishable from 0 (Lucas 2013, p. 625, Table 4).[19] The difference between first and later births is another fingerprint, evidently left by the biological consequences of eradication.

[19] In the baseline hazard regression for second births, the coefficient on a region's 1937 spleen rate×pre-1947 dummy is –0.136 (standard error 0.060). This indicates that after the eradication campaign, living in a malarial area increased the probability per unit time of a woman becoming pregnant. However, when the sample is restricted to 20 years centered around the eradication campaign, 1938–58, in which the campaign is most plausibly influential, the coefficient flips to 0.075 (standard error 0.056).

The quasi-experiment demonstrates rather convincingly that eradicating malaria increased fertility, at least by increasing the probability of successful completion of first births. Technically, that is what this review is after: rigorous evidence of the impact of life-saving interventions on fertility. However, it makes evident that conceptual line between mortality and fertility is thin. Saving a pre-term baby a month before its due date is increased fertility. Saving it a month after birth is reduced mortality. In spirit, the Lucas (2013) finding is closer to an impact on mortality than the sort of knock-on effect on fertility that we set out to find.

### 6.3 Kumar (2009), "Fertility and birth spacing consequences of childhood immunization program: Evidence from India," Ph.D. dissertation, University of Houston

Kumar (2009) looks at how fertility evolved in India after a national childhood immunization program was rolled out across the states over 1985-90. The study is constructed much like Lucas (2013), from birth histories collected through a national survey after the campaign.

Kumar (2009, p. 25), finds that a woman who started her family—had her first child—after the immunization program launched in her district had births spaced farther apart. The probability that her second birth followed within 2 years of the first fell 1.4%; within 3 years by 2.3%; and within 5 years by 1.5%. The regressions control for a woman's religion, caste, poverty level, age at marriage, and other characteristics. They include district fixed effects, which should ward off the theory that the results are explained merely by the fact that poorer districts, by virtue of being poor, had higher fertility and got the immunization program later.

However, unlike the Lucas regressions, the Kumar (2009) ones do not control for time trends, which would allow us to similarly dismiss a theory of general fertility convergence across India. Nor does it include graphs like those in Bleakley (2010) and Lucas (2013) in order to check this possibility visually. On the evidence in the paper, the correlation between the arrival of immunization and falling fertility could just reflect strong convergence on both fronts across India in the 1980s and 1990s.[20]

### 6.4 Wilson (2013), "Child Mortality Risk and Fertility: Evidence from Prevention of Mother-to-Child Transmission of HIV," revised and resubmitted to *Journal of Development Economics*

According to a source cited in Wilson (2013, p. 4), approximately a tenth of all HIV infections circa 2000 were caused by mothers transmitting the virus to their babies, whether *in utero*, during birth, or through breastfeeding. Prevention of mother-to-child transmission (PMTCT) programs arose in many countries to combat this problem. In Zambia, the site of the Wilson study, PMTCT efforts began around 2000, apparently focusing on providing HIV tests and counseling expectant mothers on such matters as breastfeeding. Within a few years, antiretroviral provision entered PMTCT practice in Zambia. "Single-dose nevirapine (NVP) was the main prophylaxis in the early years of the Zambian PMTCT program and zidovudine (ZDV) (also known as azidothymidine (AZT)) and NVP in the later years of the program" (Wilson (2013), p. 6).

Until recently, the ability of the HIV virus to spread through breastfeeding created an excruciating choice for PMTCT programs: discourage breastfeeding in order to reduce infection (25–60% of babies who got HIV from their mothers died before age 2 circa 2000 (Wilson 2013, p. 4)), or encourage it anyway because of its separate life-saving benefits. The Zambia government appears to have taken its first position on the issue in 2007, which approximated neutrality: "HIV positive mothers were to be 'given enough information about advantages and disadvantages of the available options for them to be able to make an informed choice about what might be best for them' and advised to completely avoid breastfeeding when quality replacement feeding was available" (Wilson 2013, p. 5). This ambiguity appears to have reflected the muddle in practice in the preceding years: "Evidence from Ndola city…suggests clinics may have offered advice similar to the 2007 National Protocol Guidelines. However, evidence…suggests that breastfeeding advice varied across health workers within a given clinic" (Wilson 2013, p. 5–6).

[20] Turan (2011, p. 25), documents strong convergence across Indian states in child mortality during the 1980s. The immunization program was probably a major driver.

After a randomized study in Kenya, Burkina Faso, and South Africa demonstrated that triple-antiretroviral drug prophylaxis for the mother was safe and reduced transmission through breastfeeding (Kesho Bora Study Group 2011), the government of Zambia, like the WHO (WHO 2010), revised its guidelines to favor breastfeeding in tandem with such treatment. The data for the Wilson study, however, were gathered before that, through four surveys fielded in 2001–07.

Wilson (2013) largely follows the quasi-experimental template of the studies just above. It studies an episode during which a particular health intervention was rolled out across a territory—in this case PMTCT across Zambia during 2000–07. In particular, Wilson (2013) studies whether women living within 20 kilometers of an active PMTCT site, such as a district clinic, were more likely to breastfeed or less likely to get pregnant. Wilson (2013) differs from the above studies in not reconstructing birth histories for each woman interviewed by surveyors. At least as far as this analysis goes, women were merely asked whether they had been pregnant any time in the last 12 months; and whether, if they had a child under 2, they were breastfeeding. But since PMTCT programs arrived in different places at different times, Wilson (2013) can still examine the evolution of correlations with pregnancy and breastfeeding. That is, Wilson can compute the average breastfeeding rate among women who have had have had access to PMTCT for 1 year, 2 years, 3 years, etc.

A strength of Wilson (2013) is its recognition that the rollout of PMTCT was hardly a perfect quasi-experiment. It was not random: people near the programs were better educated on average, closer to major roads, less likely to be married, and more likely to be HIV-positive (Wilson 2013, Table 2 & Fig 2). These correlations open the door to competing theories for any results found. In response, Wilson (2013) stepwise introduces an aggressive set of controls. These include individual traits such as age, years of schooling completed, and number of children. They include indicators for each month and year during which interviews took place in order to remove any long-term trends in national-average PMTCT access, breastfeeding, and fertility rates. As in Lucas (2013), the controls also include variables to remove province-level fixed effects and linear trends, such as might occur in general convergence. And they include indicators for the spread of other life-saving interventions: bed nets, piped water, and access to other HIV/AIDS prevention and treatment services.

One peculiarity in this quasi-experiment is the ambiguity of the intervention. We don't know if women were advised more for or against breastfeeding. And whether more or less breastfeeding saved or cost lives depended in part on another unknown: how many mothers were introduced to or already taking antiretrovirals, without which breastfeeding would have more likely transmitted HIV. This is a major weakness from our point of view, since Wilson (2013) does not look at child mortality, and without data it is hard to be confident even of the sign of the impact of PMTCT on child survival. When Wilson (2013) finds that access to PMTCT is correlated with lower fertility (see below), it is not clear whether to interpret that as the result of higher or lower child mortality, or neither.

Wilson (2013) finds that mothers of under-2s were 3.7–23.6% more likely to be breastfeeding if living within 20 kilometers of a PMTCT site (Wilson 2013, Table 3, Panel B). The high end of that range arises from the most conservative regression, with the fullest set of controls (Col 6). Meanwhile, mothers near PMTCT sites were 2.0–8.9% less likely to be pregnant—although here the *smallest* value comes from the most conservative regression, and is not statistically significant by conventional standards (Wilson 2013,

Table 3, Panel A).[21] Since this pregnancy regression is the one that controls for the rollout of other lifesaving interventions, we cannot, going just by these results, fully rule out the theory that the negative correlation between PMTCT access and pregnancy is caused by temporal and geographic similarities in the patterns of arrival of PMTCT and at least one of the other interventions.

Graphical analysis, however, mostly supports PMTCT as the driver. This graph (Wilson 2013, Fig 3) shows the change in the chance of a woman being pregnant relative to when a PMTCT program first began operating near her. (By construction, that change is 0 at time 0, when the program arrives.) The data include adjustments for the individual and household traits:

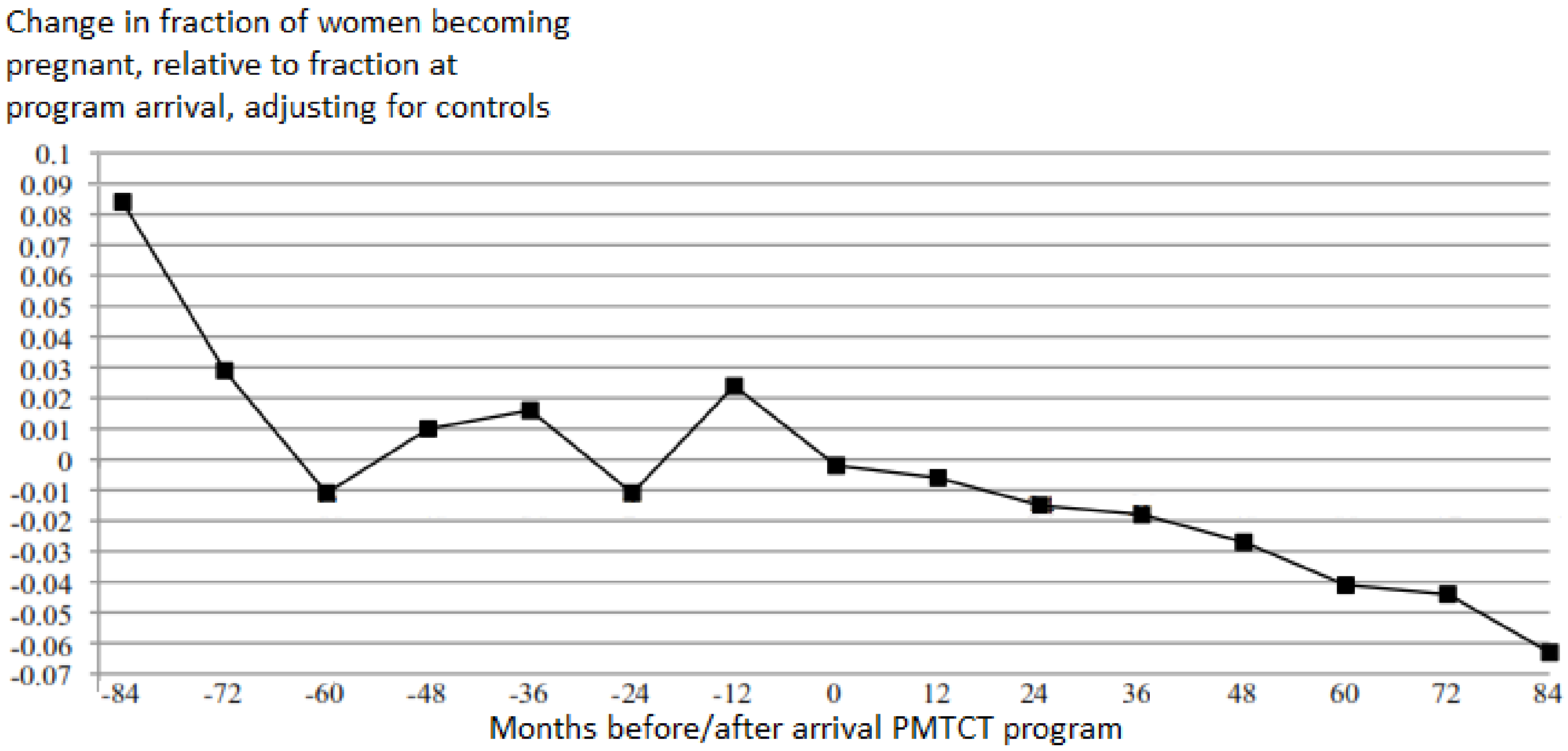

The high values at 84 and 72 months (7 and 6 years) appear to be a statistically insignificant aberration caused by very small samples of women living in areas where programs were set up so many years after the surveyor visited.[22] That aside, for the five years before PMTCT, fertility is static. After arrival—month zero—it begins to fall steadily. This favors PMTCT as the true cause of the decline.

As noted, Wilson (2013) presents no data on mortality. As a result, the simplest explanation for its results is that:

1. PMTCT programs persuaded women to breastfeed longer, for better and worse.
2. Lactational amenorrhea reduced fertility , a channel that Wilson recognizes.

This story does not require causation of fertility by mortality or even morbidity. As a result, Wilson (2013) has limited relevance for our inquiry.

[21] The point estimate is –0.02 and the standard error 0.017, for a two-tailed *p* value of 0.24. That is, if the true correlation is 0 and the statistical model correct, the probability of obtaining a coefficient so large in magnitude—less than –0.02 or greater than 0.02—is 0.24. Other regressions with full controls (Cols 6 of Tables 4, 5, 6) produce coefficients at least as weak.

[22] E-mail from Nicholas Wilson, April 1, 2014.

## 6.5 Bhalotra, Hollywood, and Venkataramani (2012), "Fertility Responses to Infant and Maternal Mortality: Quasi-Experimental Evidence from 20th Century America," preliminary and incomplete

Bhalotra, Hollywood, and Venkataramani (2012) sets out to trace the consequences of the introduction of the first (sulfanomide) antibiotics in the United States. Adoption was rapid, and the authors date it to 1937. More than in any other quasi-experimental study reviewed here, this one involves a change in the health regime with implications for many diseases, which affect both mortality and morbidity for both mother and child. The operation of multiple channels creates the risk that those captured in the analysis are merely proxying for ones left out, and that those that are captured offset or reinforce each other in ways hard to disentangle.

Bhalotra, Hollywood, and Venkataramani (2012) focus on two diseases that were plausibly the dominant channels of impact for sulfa drugs. One affects babies: infant pneumonia. "Pneumonia was the leading cause of infant morbidity and mortality after death from premature birth and congenital defects and it is similarly the leading cause of infant death in developing countries today" (Bhalotra, Hollywood, and Venkataramani 2012, p. 5). The other affects mothers, as mothers: puerperal fever, an infection contracted during birth. "Such infections remain the leading cause of maternal mortality in the developing world and, among those who survive, can cause uterine scarring and infertility." The authors demonstrate through graphs that the trend in both diseases bent downward around 1937.

Before 1937, the diseases varied in prevalence across U.S. states, and differed from each other in these patterns of prevalence. With the introduction of the drugs, the *declines* in infant and maternal mortality therefore differed by state and from each other. This allows Bhalotra, Hollywood, and Venkataramani (2012) to study the correlations of infant and maternal mortality with fertility simultaneously in the years that followed.

As usual, a primary concern is that long-term convergence—in this case, among U.S. states—in mortality and fertility will be mistaken for one causing the other. Bhalotra, Hollywood, and Venkataramani (2012) take the appropriate step of introducing controls that allow each state to have its own long-term linear trend in fertility. As in the Bleakley Brazil example, in the regressions including these controls—the most conservative ones—only deviations from straight-line trends, influence the results.

However, I believe there is a serious limitation in this preliminary study, or at least in the data it has to work with. The fingerprint of the impact of the arrival of sulfa drugs on fertility appears to be missing. The paper does not search for it graphically. And the statistical results say that the effects of reduced infant and maternal mortality on fertility largely cancel out. "A change in [a state's pre-1937 infant pneumonia prevalence] equivalent to a movement from the 75th percentile to 25th percentile...implies a 0.47% point *drop in the probability of a birth after 1937, which is 4.3% of the mean* (0.11). A similar shift in [a state's pre-1937 maternal mortality rate] implies a 0.48% point *increase in the probability of a birth after 1937, 4.4% of mean*. [Emphasis added.]" (Bhalotra, Hollywood, and Venkataramani 2012, p. 12).

The inclusion of controls for linear trends cannot solve this problem, and actually produces results that increase my doubts. It cannot solve the problem because, absent a corner in 1937, it forces the infant and maternal mortality indicators to search for other deviations from straight-line trends to explain. Convergence could well occur with curvature: as declines proceed, they could decelerate, as room for further decline shrinks. Curvature would leave some discrepancies from the linear trend to be falsely attributed to the impact of falling mortality. In fact, controlling for linear trends triples the apparent impact of infant mortality on fertility and quadruples that of maternal mortality (Table 3, cols 5–6). This suggests that the state trends are quite similar in time profile with the effects ascribed to the mortality changes, if of opposite sign (substantially, negatively collinear). Added to the mix is the partial collinearity in the pre-1937 geographic patterns of infant and maternal mortality, and thus of the subsequent drops therein. When the partially collinear variables enter regressions together, as they usually do in this study, they can receive large coefficients but nonetheless have very little collective explanatory power, as their effects are estimated to cancel out.

On balance, the premise of this study is smart, that the arrival of sulfa drugs offer a potentially valuable quasi-experiment, but the apparent lack of a sharp drop in fertility combines with the multiple disease channels to leave somewhat unclear what is driving the results.

## 6.6 Juhn, Kalemli-Ozcan, and Turan (2009), "HIV and Fertility in Africa: First Evidence from Population Based Surveys," Leibniz Information Centre for Economics discussion paper; and Fortson (2009), "HIV/AIDS and Fertility," *American Economic Journal: Applied Economics*

The last quasi-experiment to appear in this review is not a sudden improvement in the health regime, but a worsening: the spread of HIV in Sub-Saharan Africa.

Many of the studies already reviewed use data from Demographic Health Surveys, a family of household surveys in developing countries that USAID has supported for 30 years. (Some studies use the data directly while others take national averages derived from the surveys.) Timing tends to be quinquennial.[23] Recognizing the poor quality of data on HIV prevalence in developing countries, and taking advantage of cost-saving scientific advances, DHS surveys began in 2001 to include voluntary HIV testing. This has given researchers large datasets that link a person's HIV status with such traits as education, marital status, and fertility.

Juhn, Kalemli-Ozcan, and Turan (2009) and Fortson (2009) are among the first economics papers to harvest this data. The two studies substantially overlap in data and methods. Fortson (2009) uses survey data from Cameron, Côte d'Ivoire, Ethiopia, Ghana, Kenya, Malawi, Mali, Niger, Rwanda, Tanzania, Zambia, and Zimbabwe, gathered between 2001 and 2006. Juhn, Kalemli-Ozcan, and Turan (2009) drops Mali and Zambia and adds Burkina Faso, Guinea, and Senegal. Neither paper claims "quasi-experimental" status, perhaps because the spread of HIV was more gradual than a DDT spraying campaign as well as less exogenous, being modulated by local cultural, social, and economic factors. Still, the studies fit in the quasi-experimental family in analyzing the consequences of a major, novel, and relatively swift shift in the health regime.

Both studies recognize that HIV, like malaria, can affect fertility through biological and volitional channels. As a matter of biology, it can increase miscarriages and vulnerability to other infections that can cause infertility (Juhn, Kalemli-Ozcan, and Turan 2009, Pg, 5–6). As for volition, knowing that a woman is HIV+ or that HIV is spreading locally could affect a couple's preferences and decisions about having children. Recall that Lucas (2013) distinguished biological and volitional effects of malaria by looking separately at women giving birth for the first time, since they are more susceptible to the disease. HIV is not known to affect first births differently. But researchers can still distinguish the channels by separately studying the impact of the spread of HIV in a community on women who test negative in the survey (but who might not have known their HIV status before); these women are, as it were, immune to the biological channel.

In Juhn, Kalemli-Ozcan, and Turan (2009), about 85% of respondents agreed to HIV testing. Among them, being HIV+ reduced the probability that they had given birth within the year leading up to the survey by 3.4%, and by 9.2% for the last 3 years and 13.6% for the last 5 (Juhn, Kalemli-Ozcan, and Turan 2009, Table

[23] Data are at dhsprogram.com/data/available-datasets.cfm.

5, col 3). HIV seemed to lower fertility.

At least two considerations should impede us from immediately taking this result at face value. First, the women who refused the test may have been systematically more or less HIV+ on average, in which case their absence would throw the results a bit. Juhn, Kalemli-Ozcan, and Turan (2009) finds test refusers to be more educated, wealthy, and urban, making them statistically akin to respondents who tested HIV+ (Juhn, Kalemli-Ozcan, and Turan 2009, p. 8). Second, and more important, confounders compete to explain the correlation. Education, wealth, and urban location could lead to lower fertility along with higher HIV prevalence.

But Juhn, Kalemli-Ozcan, and Turan (2009) control for such demographic factors, along with indicators of sexual behavior such as condom use, and their results stand. Meanwhile, the biological channel alone credibly explains the correlation as causation from HIV to lower fertility.

To isolate biology from cognition, Juhn, Kalemli-Ozcan, and Turan (2009) next calculate whether, HIV-negative women in high-HIV regions have fewer babies than HIV-negative women elsewhere. This could indicate that *anticipation* of mortality affects fertility. It would be relevant to the question of whether couples deliberately adjust family size in response to perceived mortality risks in their community. The answer from Juhn, Kalemli-Ozcan, and Turan (2009) turns out to be "no." The paper confirms the robustness of this result in a few ways, such as by instrumenting HIV prevalence with distance from the Democratic Republic of Congo, where the virus originated (Juhn, Kalemli-Ozcan, and Turan 2009, Tables 9–11).

Fortson (2009) reaches the same conclusion independently, despite a somewhat different approach. The major methodological difference is that Fortson (2009), like most of the other quasi-experimental studies, reconstructs full birth histories for each woman surveyed. This results in a data set on fertility by subnational region (there are 108 within the 12 countries) and by year, rather than just by region. Having observations at multiple times within regions allows Fortson (2009) to remove any national trends, linear or otherwise.[24] Paralleling Juhn, Kalemli-Ozcan, and Turan (2009), Fortson( 2009, Table 3, Panel A, col 3), computes that HIV+ women have 0.146 fewer lifetime births.[25] But among HIV-negative women, those living in high-HIV regions are not detectably more or less fertile than HIV-negative women in low-HIV regions. Again, there is no support for a volitional pathway from mortality to fertility (Fortson 2009, p. 180, Table 3, Panel B).

## 6.7 Summary: quasi-experimental studies

Of the 5 studies in this section, several provide credible evidence of impacts on fertility. Lucas (2013) (on malaria in Sri Lanka) and Wilson (2013) (on PMTCT in Zambia) graph trend breaks that are rather convincingly timed. Juhn, Kalemli-Ozcan, and Turan (2009) and Fortson (2009) correlate HIV status with fertility. But none of the studies turns out to speak directly to our interest. In Sri Lanka, malaria eradication directly *raised* fertility in a way that is conceptually tantamount to lowering mortality, rather than being a knock-on effect from changed mortality. Similarly, the multi-country Africa HIV studies mostly suggest that HIV biologically reduces fertility. In Zambia, PMTCT may have reduced fertility mainly by encouraging breastfeeding.

---

[24] The preferred regressions (Table 2, col 6; Table 3, Panel B, col 4) have one observation for each region-year combination. They include country-year dummies, and region fixed effects. HIV prevalence is assumed zero through 1990, and constant at the survey values in 2000 and later. Observations for 1991–99 are excluded for lack of data on HIV prevalence.

[25] The standard deviation of 0.500 corresponds to a standard error of about 0.500/sqrt(108)=0.048 if the region is taken as the unit of observation, there being 108 regions. This puts 0.146 3 standard errors above 0, for high statistical significance.

# 7 Large-sample micro-studies

The final class of studies reviewed here consists of ones based on data sets similar to those in the previous class. Data are collected through interviews with adults in households. Birth histories of women are then reconstructed. These studies stake no claim to quasi-experimental status. But some offer an alternative virtue: very large data sets, which allow the researchers to aggressively control for variables embodying theories they want to rule out, and to use clever techniques to distinguish the various channels by which mortality affects fertility.

For a statistician, a large data set is like a powerful telescope. The more independent observations she has, be they of families, trees, or stars, the more confidently she can discern statistical relationships. Flip a coin once and you cannot tell if it is fair. Flip it a million times and you almost certainly can. With large data sets, researchers can control for many third factors without worry about depleting the statistical power needed to detect the relationships of interest, in this case between mortality and fertility. The results they obtain are still only correlations. But they are ones for which, ideally, many explanations have been ruled out, raising confidence in those that remain.

Although these studies draw on similar data to the quasi-experimental studies in the last section, they make a fundamental but subtle shift in perspective. Before, the source of identifying variation—t he basis of the quasi-experiment—was regional because diseases arose and disappeared on that scale. Now the family in not only the unit of observation but the source of variation. The natural focus now is on whether a death *in the family* changes reproductive behavior in that family. Effects of regional changes in the health regime—a family having fewer kids because it perceives that fewer children are dying in its community—are harder to capture. But the studies below try.

Hundreds of representative households surveys have been conducted in developing countries in recent decades, and many of them obtained birth and death histories from respondents. As a result, many mortality-fertility studies have been based on them. Here I review some of the more recent and rigorous ones.

## 7.1 Bhalotra and van Soest (2008), "Birth-spacing, Fertility and Neonatal Mortality in India: Dynamics, Frailty, and Fecundity," *Journal of Econometrics*

The most complex study in this group is Bhalotra and van Soest (2008), which works on data about 30,000 births between 1963 and 1999 in 7,300 households surveyed in Uttar Pradesh state, India, in 1998–99. Uttar Pradesh is one of the poorest states in India, and in one 2004 survey had the country's highest fertility rate, at 4.39 births/woman (Haub 2011, p. 21). Still that was down from about 6.5 in 1972, suggesting that family planning was spreading during the years covered by the women's birth histories.

The regressions are rather like the Granger-inspired ones in Herzer, Strulik, and Vollmer (2012). The three outcomes of greatest interest are whether a given child dies in its first month of life; whether, either way, the mother has another child after this child; and, if she does, how many months pass between births. The first two variables, mortality and birth interval, can interact over time. If a child dies, its mother may get pregnant again sooner. If she gets pregnant sooner, the next child will face tougher survival odds (Rutstein and Winter 2014, p. 38, Table 13). Research points to several reasons: soon after one pregnancy, the mother's body may be less fit for the next; and greater competition among closely spaced siblings may reduce the sustenance and attention that each receives (Rutstein and Winter 2014, p. 1).

The regressions include many variables. Whether an infant survives its first month is allowed to depend on the survival outcomes and birth intervals for all of the mother's previous children, and vice versa. Also controlled for is a household's caste (scheduled caste, scheduled tribe, other backward caste, other, or unknown) and religion (Hindu, Muslim, or other), mother's and father's education levels, the child's gender, year of birth, and birth order, and the mother's age. All controls that can take a continuous range of values are included in ways that flexibly allow for nonlinear (curved) relationships between them and the three outcome variables. Education levels are quantized into five categories, each allowed to have a different impact. The birth year, birth order, and mother's age enter quadratically, so that they can exhibit "U"-shaped relationships with the outcomes. And they allow for fixed effects for each mother and each community surveyed: some mothers and some communities might experience higher fertility and mortality across their lifetimes because of third factors not picked up in the variable list above.[26]

Bhalotra and van Soest (2008) focus on *neonatal* mortality—death within the first month—in order to simplify their statistical model in one respect and make it more practical for fitting to actual data. If they had studied mortality over 1 or 5 years, they would have had to allow for the potential impact of *later* births on a given child's survival. If a child gained a younger sibling at 2 and died at 4, then the later arrival might have affected the fate of the earlier one. But survival for the first month can be assumed to be influenced only by earlier births, making the math simpler.

After fitting their regression model to the data, Bhalotra and van Soest (2008, Table 3), perform simulations to interpret it. For each mother in the data, the regression results are used to estimate the probability that her first child would survive its first month, considering the mother's age, caste, etc. As an example, the probability could be 90%. A random number is drawn between 0% and 100%. If it is below 90%, the child survives in the simulation. Then, the computer simulates whether the mother becomes pregnant again and, if so, after how many months. The process repeats until the simulated mother does not get pregnant again. Each step is influenced by the mother's full history of births and deaths to that point.

Bhalotra and van Soest (2008) also run the simulations with some modifications: they zero out some or all of the links their model allows between mortality and fertility. This lets them partly distinguish the channels that link mortality and fertility, assuming their model is accurate. The first column of the table below assumes neonatal mortality has *no* impact on fertility. In this alternate-universe Uttar Pradesh, women go 31.39 months between births, and 0.26 of their 3.99 children die in the first month of life.

If the community- and mother-level fixed effects are accounted for—by which higher mortality in a family or community correlates across all births with higher fertility, for reasons that are not really explained by the model—then months between births falls slightly to 31.09 and number of births and neonatal deaths rise slightly to 4.02 and 0.27 (second column). One explanation for the slight rise in fertility is that families that experience or witness more neonatal deaths have more kids, and have them sooner, to compensate for the expectation that more will die ("hoarding"). However, other explanations are available: fertility and mortality may go hand-in-hand because of badly run clinics. In turn, infant mortality can rise both because more babies are born and because they are born closer together.

The third column introduces time's arrow. It allows the fate of the *last* baby to predict that of the *next*. Why might the death of one baby make life riskier for the next? One reason, Bhalotra and van Soest (2008) explains, is that when a baby dies right after birth, the mother is more likely to have another soon. Tighter birth spacing is known to be associated with higher risk of death for subsequent children. At any rate, allowing for this link hardly changes the number of children born/family.

The final two columns allow for the finding that a mother whose infant has died is more likely to get pregnant again, and sooner. This is called the "replacement effect," on the idea that parents are having more children to replace the ones they lost. On average, simulated women now go 30.59 months between births, and have 4.13 children, of whom 0.31 (7.4%) die in the first month. Since these figures incorporate all the effects modeled, they are closest to the true values in the survey.

[26] However, with 24.4 mothers/cluster and 4.1 births/mother, Bhalotra and van Soest (2008) judged the statistical power inadequate for independent fixed effects for each mother. They impose the assumption that the 333 community-level impact values and 7,286 mother-level impact values come from a normal distribution, so they estimate only the parameters of this distribution not the individual draws from it, in a "multilevel random effects" model.

**TABLE 1. IMPACTS OF NEONATAL DEATH RATE ON SIMULATED FERTILITY AND MORTALITY IN UTTAR PRADESH, BHALOTRA AND VAN SOEST (2008)**

| | Causal channels modeled | | | | |
|---|---|---|---|---|---|
| | None | Association across communities between higher fertility and higher mortality | + Last baby dying raises death risk for next | + Last baby dying leads to higher chance of having another | + Last baby dying leads to having next sooner |
| Months between births | 31.39 | 31.09 | 31.09 | 31.11 | 30.59 |
| Number of children born/family | 3.99 | 4.02 | 4.02 | 4.10 | 4.13 |
| Number dying <1 month | 0.26 | 0.27 | 0.29 | 0.29 | 0.31 |
| Number surviving >1 month | 3.72 | 3.75 | 3.73 | 3.80 | 3.82 |

Note: Causal channels are factored in cumulatively, moving from left to right. Last column factors in all statistical relationships in the underlying model. Since all the numbers are derived from this full model after it has been calibrated to the data, the last column best matches the observed averages.

Moving from the first to the last column, as all the statistical ripple effects of neonatal mortality are added in, the number of children born per family climbs by 4.13–3.99 = 0.14. If we fully attribute this fertility increase to ongoing mortality—0.26 deaths/family in the base case of the first column—we conclude that each neonatal death caused women to have 0.14/0.26 = 0.52 more children. Since hoarding is not the only viable explanation for the fertility climb between columns 1 and 3, we can conservatively count just the replacement effect manifesting from column 3 to column 5. (Here, the effects are Granger-style, with earlier events allowed to affect later ones, making the causal chain from mortality to fertility more convincing.) The replacement effect is 4.13 – 4.01 = 0.11 extra children per family. Blame that on the deaths of 0.29 neonates/family listed in column 3, and we concluded that families go on to have 0.11/0.29 = 0.37 more babies for each they lose in the first month of life.

While we cannot ascribe causal significance to this result as confidently as in a clean experiment, it does have time's arrow in its favor. Possibly the true effect is much larger and some countervailing causal story is counteracting it, such as disease reduction simultaneously reducing deaths and increasing fertility. On the other hand, the calculations above probably overstate the effect in a different way. For they take the ratio of *all* extra births to a subset of the deaths, just neonatal ones. If a high neonatal death rate is proxying for higher under-5 and under-10 death rates, then the true mortality-change-fertility-change causal ratio might be only half the 0.37–0.52 suggested above.

On balance these results make a true average effect as large as 1:1 seem unlikely in Uttar Pradesh across 1963–99. This suggests that saving lives in Uttar Pradesh accelerated population growth on average.

## 7.2 Hossain, Phillips, and LeGrand (2007), "The Impact of Childhood Mortality on Fertility in Six Rural Thanas of Bangladesh," *Demography*

The data set for this study is distinctive in coming from longitudinal surveillance. The source is not a single, randomized survey from which birth and death histories are constructed (based on potentially faulty recollections of events long past). Nor is it a series of randomized surveys in a given country, each interviewing different people. Rather, 8,000 women in the Matlab district of Bangladesh were randomly chosen in 1982, and repeatedly visited through 1993, to gather recollections of household events when they were fresher.

Those years proved a remarkable time in Bangladesh. Nationally, total fertility fell from 6.5 to 3.5 births/woman between 1980 and 1995. "The pace of reproductive change in this period ranks among the most rapid ever recorded" (Hossain, Phillips, and LeGrand 2007, p. 773).

Hossain, Phillips, and LeGrand (2007) differs greatly from Bhalotra and van Soest (2008) in method. It starts with the observation that how the death of child within the family is correlated with or causally connected to the decision to have another depends on who dies and when. Imagine a women who has just given birth, and ask: on average, how long will it be before she becomes pregnant again, and how does that depend on the family's history of child deaths? Borrowing from work of other authors, Hossain, Phillips, and LeGrand (2007) enumerate these possibilities:

- If having another child sooner after this birth is correlated with having lost a child *before* this recent birth, that would mainly point to a forward-looking "hoarding" effect: a history of loss leads to anticipation of future loss, which leads to long-term desire for extra children to compensate.
- If having another child sooner after this birth is correlated with having lost an older child *after* this recent birth, that suggests a backward-looking, volitional *replacement* effect.
- If having another child sooner after this birth is correlated with the just-born child dying, that would pick up both the volitional replacement motive above and biological replacement mechanism of interrupted lactational amenorrhea. We could compare the size of this correlation to the previous one to estimate the additional impact of the biological pathway.
- If having another child sooner after this birth is correlated with losing a child well in the *future*, since the future death couldn't cause the decision to get pregnant sooner now, this would suggest that ongoing third factors such as poverty are causing the family to have more births and deaths. Including future deaths in the regressions becomes a way to control for such third variables.

Hossain, Phillips, and LeGrand (2007, p. 778, Table 2, col 3), finds strong positive associations in all cases. If the given newborn soon dies, that multiplies the probability per unit time that the mother will get pregnant again by 59.7—from about 1 in 300, for example, to 1 in 5.[27] That is the combination of the biological and volitional replacement effects. The death of an older child, which would trigger only the volitional replacement effect since the mother could still breastfeed, expands the probability by "only" a factor of 6.2. The death of a child farther in the past, indicating a history of loss and expectation of future loss, has a smaller effect still, at 3.5.

| Event | Multiplier on probability per month of getting pregnant again |
|---|---|
| Child died before most recent birth (hoarding) | 3.5 |
| Older child died after most recent birth (volitional replacement) | 6.2 |
| Newborn died (volitional + biological replacement) | 59.7 |
| A child dies in year *after* next pregnancy (third factors raising births and deaths) | 9.7 |

Unfortunately, Hossain, Phillips, and LeGrand (2007) does not translate these abstract results into a form that compares total births to total deaths. So it is hard to infer implications for the impact of saving lives on population growth. The results do suggest that the short-term biological link from death to birth is large

[27] These results are from the regression with the largest control set. The dependent variable is the log of the hazard ratio, which is the log of the probability per unit time that an event will occur if it has not yet occurred. The tabulated numbers are antilogarithms of coefficient estimates in Hossain, Phillips, and LeGrand (2007).

enough to statistically mask the volitional effect unless controlled for.

## 7.3 Binka, Bawah, and Hossain (2004), "The Role of Childhood Mortality in Fertility Transition in a Rural Sahelian District of Northern Ghana," incomplete

The Navrongo district in northern Ghana, like Matlab in Bangladesh, has been the site of extensive demographic data collection over the years, which has facilitated research on family planning in the extremely poor Sahelian region of Africa. Binka, Bawah, and Hossain (2004) was presented at a conference in 2004 and was posted as an incomplete 3-page document. The results, based on data from an impressive 43,000 women, are that "the death of a child has no effect on the odds of subsequent parity progression," i.e., having another child. This is tantalizing evidence that in the poorest places, where families have not yet begun to limit fertility, the loss of a child matters little for whether the woman gets pregnant again. If she does get pregnant again, she would have anyway. The result makes sense, but unfortunately cannot be evaluated without a full paper.

## 7.4 Older microdata studies

As mentioned, there are many studies in this genre. This quote from the Schultz (1997), Pg. 384–85, review suggests that the older ones line up with Bhalotra and van Soest (2008) in finding a mortality-drop-fertility-drop ratio of less than 1:1.

> Using the 1973 Census sample of Colombia, Olsen (1980) estimated that the replacement response effect was about 0.3, rather than the ordinary least squares [no-instrument] estimate of 0.5, suggesting that for every three child deaths prevented there was one fewer birth. Rosenzweig and Schultz (1982b) estimated, from the same data source using instrumental variables method, the sum of replacement and expectation response rates of between 0.14 and 0.42 for various cohorts of women between the age of 24 and 54. Lee and Schultz (1982) estimated the replacement response in Korea in 1971 as between 0.35 and 0.51….Maglad (1990), using the instrumental variables method, estimated for a small sample in rural Sudan a replacement/expectation ["hoarding"] rate in 1987 of between 0.56 and 0.73. Okojie (1991) obtained significant estimates of replacement/expectation responses for Bendel state of Nigeria from a sample collected in 1985. Benefo and Schultz (1992) estimated by instrumental variables a replacement/expectation rate of about 0.2 from a national sample of Ghana collected in 1987-1989 and obtained a similar value for Côte d'Ivoire in 1985–1988. Mauskopf and Wallace (1984) estimated the replacement probability for Brazil was nearly 0.6 and found that it increased from 0.44 to 0.98 as the woman's education increased from none to five or more years….Finally, in a high-income environment, Rosenzweig and Schultz (1983) estimated by instrumental variables a replacement effect of about 0.2 from a 1967–1969 sample of legitimate births for the United States.

Not having reviewed these studies, I cannot comment on the strength of their conclusions. I suspect that to delve into them would be to reap diminishing returns.

# 8 Conclusion

Table 2 summarizes my current interpretations of the studies reviewed above:

None of the studies in this review produces evidence that is both relevant to our question and beyond challenge. However, if we combine the best evidence with general knowledge about the spread of family planning, a consistent picture emerges.

In my view,

- The message of the historical evidence is reasonable: the long delays between mortality declines

and the onsets of fertility declines suggest that there is more behind the latter than the former.

- The cross-country studies seem most suspect. They imply very large impacts of mortality decline on fertility. I find it hard to reject the hypothesis that their results are driven by the crude 1:4 ratio revealed in the cross-country graph for 1990 in the "Mortality and fertility: Trends and causes" section above.
- Some of the quasi-experimental studies produce convincing results. The closest they come to answering our empirical question is in suggesting that health interventions for *women of child-bearing age* make them more fertile, e.g., by helping them bring pregnancies to term. This is worth noting, especially since the effect is opposite in sign of what we were looking for: life-saving interventions reduce deaths and *increase* births. It would not generalize to interventions aimed at other demographics, such as children.
- I find the Granger-style studies, the ones that systematically explore the relationships over time between deaths and births, most useful. They do not aspire to measure true causality—only what happens after what, on average. But if that aspiration of measuring true causality is unrealistic, then perhaps the humility is for the best. Working with country-level statistics, mostly from relatively wealthy countries over the 20th century, Herzer, Strulik, and Vollmer (2012) finds that drops in mortality are followed over a generation by fertility drops nearly as large. Looking within families in Uttar Pradesh in the decades up to 1999, a context in which fertility was high but had begun to fall, Bhalotra and van Soest (2008) finds partial replacement, with 0.37–0.52 extra births for each neonatal death. The incomplete Binka, Bawah, and Hossain (2004) hints that in a region where fertility was high and mostly not controlled, the loss of a child did not lead on average to a family having more thereafter.

As mentioned at the outset, we should expect that where fertility is most controlled, typically indicated by total fertility of about 2 births/woman or less, that the volitional replacement effect is large—that for every child's life saved, parents avert one birth. That births/woman averaged 2.7 in developing countries as a whole in 2005–10, and that the number has probably fallen more since, suggest that most couples today are engaging in family planning. Meanwhile, where the fertility transition does not yet appear to have occurred the replacement effect is likely much smaller. The studies I find most informative tend to corroborate this theory, indicating near-full replacement among a group of relatively affluent countries; partial replacement in a context where fertility had begun to decline but still had far to go (Uttar Pradesh); and no replacement in an area of continuing high fertility (Northern Ghana).

A corollary to this interpretation is that the mortality decline in developing countries during the last 60 years probably caused a minority of the contemporaneous fertility decline. Recall that under-15 deaths fell from 1.8 to 0.21 per woman between 1950–55 and 2005–10, a drop of 1.6. If the impact of mortality declines on fertility ranged by place and time between 1:0 and 1:1 then it could not have caused more than 1.6 of the decline in births/woman, which was from 6.1 to 2.7, a drop of 3.4 children/woman. Other factors, such as female education, economic growth, contraceptive availability and family planning promotion together likely mattered more.

### Table 2. Summary of studies and findings

| Study | Methodology & setting | Effect size (study's unit) | Effect size (births/death) | Comments |
|---|---|---|---|---|
| Schultz (1997) | Cross-country, 80 developing countries, 1972–89 | 0.25 births/woman per under-5 deaths/100 births | 5 | Interpreting cross-country correlations as influence of mortality on fertility requires strong, debatable |

| Study | Methodology & setting | Effect size (study's unit) | Effect size (births/death) | Comments |
|---|---|---|---|---|
| Conley, McCord, and Sachs (2007) | Cross-country, 138 developing countries, 1960–2004 | 0.1 births/woman per under-1 deaths/100 births | 2 | assumptions |
| Lorentzen, McMillan, and Wacziarg (2008) | Cross-country, 85 countries, 1960–2000 | 0.15 births/woman per under-1 deaths/100 births | 3 | |
| Murtin (2012) | Cross-country panel, 70 countries, 1870–2000 | 0.2–0.4% change in births/population per 1% change in under-one deaths/100 births | 2–4 | Study of changes over time within countries is closer to a controlled experiment than cross-country comparisons, since many national traits evolve only slowly. But strong assumptions still required, since as fertility and mortality change over decades, so do many other factors that influence them. |
| Herzer, Strulik, and Vollmer (2012) | Cross-country panel, 20 countries, 1900–2000 | 0.8 births/population per all-age deaths/population, long-term | 0.8, over a generation | Ditto; however, this study has virtue of transparency, in describing how fertility evolves after a change in mortality without making deeper claims as to causality |
| Lucas (2013) | Quasi-experiment based on late-1940s near-eradication of malaria, Sri Lankan households, 1939–75 | Per pre-1937 % of children with enlarged spleens (indicating malaria prevalence), % of women having live birth per year after campaign *up* 22%; % of first-borns dying before age 5 down 45%; death rate of later-born unchanged | Negative; size hard to infer because relationship between birth probability and total births is complex | Causal relationships credibly discerned, thanks to clear fertility jump after eradication. Best explanation for result is not deaths influencing births, but eradication of malaria increasing survival before and after birth. |
| Kumar (2009) | Quasi-experiment based on 1980s child immunization campaign; Indian households, ~1973–2003 | Women who had first child after the immunization program arrived in district spaced births farther apart: probability of 2nd birth within 2 years of the first fell 1.4%; within 3 years, 2.3%; and within 5 years, 1.5% | Not estimable: intervention's correlation with mortality not analyzed | Causal relationships not as credibly discerned: does not control for long-terms trends in health & fertility, such as convergence across regions; does not identify quasi-experimental discontinuities. |

| Study | Methodology & setting | Effect size (study's unit) | Effect size (births/death) | Comments |
|---|---|---|---|---|
| Wilson (2013) | Quasi-experiment based on 2000s Prevention of Mother-to-Child Transmission (PMTCT) program rollout; Zambian households, 2000–07 | Mothers near a PMTCT site 2% less likely to have been pregnant in given year | Not estimable: intervention's correlation with mortality not analyzed | Causal relationships credibly discerned, thanks to clear fertility drop after PMTCT arrival. Best explanation is: the program encouraged breastfeeding, delaying pregnancy via lactational amenorrhea—rather than falling mortality reducing fertility. |
| Bhalotra, Hollywood, and Venkataramani (2012) | Quasi-experiment based on introduction of first antibiotics in U.S. circa 1937; U.S. states, 1930–70 | Safer birth leads to more births while higher infant survival leads to fewer, to offsetting degrees | 0 (for maternal and infant death reductions altogether) | Clear drops in mortality starting circa 1937 bolster case for quasi-experiment. But results sensitive to controlling for long-terms trends in health & fertility, such as convergence among states; decomposition of impacts into offsetting maternal and infant death channels less credible than net (zero) effect. |
| Juhn, Kalemli-Ozcan, and Turan (2009) | Quasi-experiment based on spread of HIV in 13 African nations in 2000s, 2002–06 | HIV+ women 3.4% less likely to have been pregnant in last year, 9.2% in 3 years, 13.6% in last 5; among HIV-negative women, being in a high-prevalence area did not raise probability of being pregnant | Not estimable: HIV's correlation with mortality not analyzed | Interpretation of results as impact of HIV on fertility is reasonable. But best explanation is biological—HIV and associated infections impeding successful pregnancy—rather than volitional, since among HIV-negative women, *witnessing* higher mortality locally did not affect fertility. |
| Fortson (2009) | Quasi-experiment based on spread of HIV in 12 African nations in 2000s, 1981–2005 | HIV+ women have 0.15 fewer lifetime births | Not estimable: HIV's correlation with mortality not analyzed | |
| Bhalotra and van Soest (2008) | Non-experimental study based on household data for 1963–99, Uttar Pradesh, India | Death of a neonate (under 1 month) is followed in a family by 0.37–0.52 extra births | 0.37–0.52 | Ascription of causal link not as strong as in good quasi-experiments, but study is transparent like Herzer, Strulik, and Vollmer (2012), estimating how many extra births follow a death on average. |

| Study | Methodology & setting | Effect size (study's unit) | Effect size (births/death) | Comments |
|---|---|---|---|---|
| Hossain, Phillips, and LeGrand (2007) | Non-experimental study based on household data for 1982–93, Matlab, Bangladesh | An *older* sibling dying in family after a recent birth multiplies chance per unit time of getting pregnant again by 6.2. | Hard to infer from available data | Data drawn from repeated visits to same women, not imperfectly recalled birth histories. Teases apart biological and volitional effects by distinguishing by when a child death occurred in family relative to given birth. |
| Binka, Bawah, and Hossain (2004) | Non-experimental study based on household data for 1993–2003, Navrongo, Ghana | Child death in family does not raise chance of having another. | 0 | Study incomplete. Results only suggestive. |